\documentclass[12pt]{iopart}

\usepackage{graphicx}
\usepackage[below]{placeins}

\newcommand{\bx}{{\bf x}}

\begin{document}

\title[CMB spectrum for binary polyhedral spaces]
{Predicting the CMB power spectrum for binary polyhedral spaces}

\author{Jesper Gundermann}
\address{Danish Environmental Protection Agency, Strandgade 29, DK-1401 K}

\begin{abstract}
The COBE and the first-year WMAP data both find the CMB quadrupole
and octopole to be anomalously low. Here it is shown, that a
finite, multi-connected universe may explain this anomaly,
supporting earlier analyses \cite{Nature} \cite{aurich2}. A novel
technique, pioneered by \cite{marc rey} is used to compute the
spectrum and its variance up to k=102. Based on the properties of
the Lie group of rotations of $S^3$ it is shown that the spectrum
and its variance may be computed solely from the matrix elements
of the group-averaging operator, for each of the manifolds
$S^3/I^*$, $S^3/O^*$ and $S^3/T^*$. Further, it is proved that the
spectrum may be calculated solely from the radial function, due to
the symmetry properties of the Lie-algebra, which is rigorously
proven. It is shown, that if the topology of the universe is
$S^3/I^*$ the uncertainty on the estimates for $\Omega_{tot}$ may
be improved by an order of magnitude. Finally, the paper
highlights how the unavailability of an explicit probability
function for the observations, given the model, is a challenge for
Monte-Carlo simulations of the spaces $S^3/\Gamma$ which has to be
addressed in future work.

\end{abstract}

\ead{jgu@mst.dk}



\section{Introduction}

Both the COBE data and the first-year WMAP data find the CMB
quadrupole and octopole to be anomalously low
\cite{hin96,bennett}.  A finite, multi-connected universe may
explain this anomaly;  see \cite{levin} for a review.  While
studies of flat spaces such as the 3-torus have not provided a
good fit to the observed CMB power spectrum, a preliminary study
of the so-called binary polyhedral spaces\footnote{See
\cite{Notices} for an elementary introduction to spherical spaces
$S^3/\Gamma$, including the binary tetrahedral, binary octahedral
and binary icosahedral spaces.} showed an excellent fit,
particularly for the dodecahedral space \cite{Nature}.
Unfortunately that preliminary study suffered two major
weaknesses:  (1)  because of computational limitations it computed
only the first two terms $C_2$ and $C_3$ of the power spectrum,
and (2) it neglected to compute the variances of the predicted
$C_i$, making a proper statistical analysis of the results
impossible. The present article resolves both those problems.  A
novel approach to the computation yields reliable predictions for
the power spectrum from $C_2$ through $C_{15}$, along with their
respective variances.

For each binary polyhedral space $S^3/\Gamma$, the computation
finds the eigenmodes of the Laplacian on $S^3$ that are invariant
under the action of the group $\Gamma$.  These $\Gamma$-invariant
eigenmodes define the basic modes of the primordial density
fluctuations, before the decoupling of the CMB.  Exploiting the
symmetry properties of these eigenmodes, expressions
for the $C_\ell$ and the variance of the $C_\ell$ are derived, that can be
computed solely from the radial functions and the matrix elements
$\langle~k \ell m~|~G~|~k \ell' m'~\rangle$ of the group-averaging
operator $G$, which projects the space of eigenfunctions on $S^3$
down to the eigenspace of $S^3/\Gamma$.  For the binary polyhedral
spaces $S^3/T^*$, $S^3/O^*$ and $S^3/I^*$, the author has carried
out the computations up to $k = 100$ in the space of 3-dimensional
modes, yielding the estimates for the $C_\ell$ and their variances
up to $\ell = 15$.

The ultimate goal of this work is, of course, to compare
predictions to observations.  Here two issues arise.  The first is
the question of what spectrum to use for the initial density
fluctuations.  The present study adopts the neutral choice of a
scale-invariant initial spectrum, keeping in mind that the
justification for such a choice is tenuous in a small universe and
may be subject to future revision.  The second issue is the
reliability of the observed low-$\ell$ CMB.  Several authors have
found unsettling alignments between the low-$\ell$ modes and the
ecliptic plane \cite{Schwarz,Hansen}, suggesting either some
hitherto unknown solar system effect on the CMB or perhaps some
errors in the collection and analysis of the data.  In light of
these uncertainties -- and the significant revisions they may
force onto the low-$\ell$ power spectrum -- it seems prudent to
wait for the second-year WMAP data as well as a plausible
resolution of the strange solar system alignments before drawing
any conclusions about which topologies best fit observations.

\section{Methods}

The space-time metric for the universe in case of both the $S^3$
and the $S^3/\Gamma$ models are given by the line-element
\begin{equation} \label{eq:m0}
  ds^2 = c^2 dt^2 - R(t)^2 dr^2 = {a(\eta)}^2 ({d\eta}^2-dr^2)
\end{equation}
where
\begin{equation}
  dr^2 = dw^2+dx^2+dy^2+dz^2
\end{equation}
is the spatial distance on the 3-dimensional unit sphere $S^3$,
$w^2+x^2+y^2+z^2=1$, and $R(t)=a(\eta)$ is the cosmic scale factor
as a function of time respectively conformal time $\eta$.

The evolution of the scale factor is given by the matter dominated
Friedmann-equation (see \cite{lehoucq})
\begin{equation} \label{eq:friedmann}
  {\left(\frac{\mathcal{H}}{\mathcal{H}_0}\right)}^2
  =\Omega_{mass}x^{-1}+\Omega_{\Lambda} x^2 - \Omega_{tot}+1
\end{equation}
where $\mathcal{H}=a'/a$, $x=a/a_0$ is the ratio of the scale
factor to the scale factor today and
$\mathcal{H}_0=|\Omega_{tot}-1|^{-1/2}$. Here $a'$ means the
derivative of $a$ with respect to the conformal time, and
$\Omega_{tot}=\Omega_{mass}+\Omega_{\Lambda}$ is the value of the
total energy-density today, expressed relative to the critical
density needed to close the universe, which has contributions from
both matter (baryons and cold dark matter) and from the
cosmological term.
\\

According to the current understanding, the observed variations in
the CMB temperatures arises as a consequence of background
fluctuations in the gravitational field, which arose during the
inflationary phase, coupled to density variations in the early
universe. After the universe became transparent to electromagnetic
radiation, the radiation coming to the observer from the last
scattering surface, will have a fingerprint from the gravitational
field, as well its value at the point from where it was originally
emitted, as from the value of the (changing) field, during its
travel to the observer. Radiation originating at a low
gravitational potential $\Phi$, will be more redshifted from
climbing out of the lower potential, and hence be seen as cooler,
than radiation originating from a local area with a high
potential. And radiation experiencing a larger than average
dilation effect from expansion during its travel to the observer
will likewise be seen as more red-shifted than the average.
 The first effect
is known as the "Sachs-Wolfe effect", the second as the
"integrated Sachs Wolfe effect". Following \cite{riazuelo} we get
the following temperature fluctuations:
\begin{equation} \label{eq:m1}
  \delta T(n)=\frac{1}{3}
   \Phi[\eta_{LSS},(\eta_0-\eta_{LSS})n]+2 \int_{\eta_{LSS}}^{\eta_0}
  \frac{\partial \Phi[\eta,(\eta_0-\eta)n]}{\partial \eta} d\eta
\end{equation}
This equation does not include a third effect, the Doppler-effect,
which arises from the relative movement towards the observer,
of the matter in the region of last scattering.
\\

The gravitational field $\Phi$ here, has its origin in the
following ansatz for a perturbed space-time metrics:
\begin{equation} \label{eq:m2}
  ds^2= a^2(\eta)[-(1+2\Phi)d \eta^2 + (1-2 \Psi)\gamma_{ij}dx^i dx^j]
\end{equation}
which is the simplest possible perturbation considering only
scalar perturbations, and working in the longitudinal gauge. On
large angular scales, one may ignore the effects of the
anisotropic pressures generated by the metric perturbations, and
find that $\Psi=\Phi$. Likewise, the influence of radiation on the
metric between the last scattering surface and today may be
neglected. Under these circumstances, the evolution of $\Phi$ is
determined by
\begin{equation} \label{eq:m3}
  \Phi''+3 \mathcal{H} (1+c_s^2)\Phi'-c_s^2 \nabla^2 \Phi+[2 \mathcal{H}' +(1+3c_s^2)({\mathcal{H}}^2-1)]\Phi=0
\end{equation}
The terms with the sound velocity $c_s^2$ may be neglected on
large angular scales, in which case we can solve the equation by
the following separation ansatz:
\begin{equation} \label{eq:m4}
  \Phi(\eta,x)= F(\eta) \sum_{\beta,s} \Phi_{\beta,s}\Psi_{\beta,s}^{|\Gamma|}(x)
\end{equation}
where $\Psi_{\beta,s}^{|\Gamma|}(x)$ are the eigenmodes of the
Laplacian compatible with the topology $S^3/\Gamma$ of the
universe, labelled with the index $\beta$, indicating the
eigenvalue $-(\beta^2-1)$ of the Laplacian, and an arbitrary
auxiliary numbering $s$ of the eigenstates. The effect of the
above various approximations is, that we can ignore the wavenumber
dependency of the perturbations, which reduce the dimensionality
of the computational problem a lot. It is however known, that a
wavenumber dependence of about 5\% is found in typical models of a
nearly flat space-time in the wavenumber range employed in this
paper (up to $\beta=101$) \cite{aurich1}. With the ansatz
(\ref{eq:m4}), we find that $F$ satisfies \cite{riazuelo}:
\begin{equation} \label{eq:m5}
  F''+3 \mathcal{H} F' + (2 \mathcal{H}' + {\mathcal{H}}^2 -1) F = 0
\end{equation}
where the derivatives are with respect to the conformal time $\eta$.
\\

In the case where the primordial fluctuations are assumed to be
adiabatic, as predicted from certain inflationary theories, the
primordial field fluctuations are assumed to be distributed with
Gaussian random complex amplitudes, on each of the eigenmodes of
the Laplacian which are allowed by the topology of the universe,
and with a power spectrum taken to be scale invariant (equal power
per logarithmic wavenumber interval), i.e. $n_s=1$, or almost
scale invariant, $n_s \simeq 1$
\begin{equation} \label{eq:m6}
  \Phi_{\beta,s} = \sqrt{P_\Phi(\beta)} X_{\beta,s} \qquad P_\Phi(\beta)
  = \frac{\alpha_P}{\beta^{n_s}(\beta^2-1)}
\end{equation}
Here $\alpha_P$ is the overall scale of the fluctuations, whereas
$X_{\beta,s}$ are random complex Gaussian variables with ensemble averages
\begin{equation}  \label{eq:m7}
  \langle X_{\beta,s} \overline{X_{\beta',s'}} \rangle
  = \delta_{\beta \beta'}\delta_{s s'}
  \qquad \langle X_{\beta,s} \rangle = \langle \overline{X_{\beta,s}}
  \overline{X_{\beta',s'}} \rangle= \langle X_{\beta,s} X_{\beta',s'} \rangle=0
\end{equation}
Third order moments have vanishing ensemble averages, while the
only nonvanishing fourth order moment is
\begin{equation}  \label{eq:m8}
  \langle X_{\beta_1,s_1}X_{\beta_2,s_2} \overline{X_{\beta_3,s_3}}
    \overline{X_{\beta_4,s_4}} \rangle
  = \delta_{\beta_1 \beta_3}\delta_{s_1 s_3}\delta_{\beta_2 \beta_4}\delta_{s_2 s_4}
  +\delta_{\beta_1 \beta_4}\delta_{s_1 s_4}\delta_{\beta_2 \beta_3}\delta_{s_2 s_3}
\end{equation}
We further expand $\langle x | \beta s \rangle =
\Psi_{\beta,s}^{|\Gamma|}(x)$ on the eigenfunctions to the
Laplacian on $S^3$
\begin{equation}  \label{eq:m9}
 \langle x | \beta \ell m \rangle
  = \Psi_{\beta\ell m}^{S^3}(x) = R_{\beta \ell}(\eta)Y_{\ell m}(\theta , \phi)
\end{equation}
\begin{equation} \label{eq:m10}
 \langle x | \beta s \rangle
 = \sum_{\ell m} \langle x | \beta \ell m \rangle \langle \beta \ell m | \beta s \rangle
\end{equation}
and find then from (1) the following expression for the temperature fluctuations
\begin{equation} \label{eq:m11}
\delta T(n)= \sum_{\beta s \ell m} K_{\beta \ell} Y_{\ell m}(\theta,\phi)
 \langle \beta \ell m| \beta s \rangle X_{\beta,s}
\end{equation}
where
\begin{equation} \label{eq:m12}
 K_{\beta \ell}
  = \left(\frac{1}{3} R_{\beta \ell}(\eta_{LSS})+2 \int_{\eta_{LSS}}^{\eta_0}
  F'(\eta)  R_{\beta \ell}(\eta_0 - \eta) d\eta \right)\sqrt{P_\Phi(\beta)}
\end{equation}
By expanding on complex basis functions and using complex random
variables, we have arrived at a result which gives a complex value
for the temperature fluctuations. The true fields and temperatures
is off course the real part
\begin{equation}  \label{eq:m13}
  2 \delta T(\theta,\phi) = \sum_{\beta s} \sum_{\ell m} 2 \Re (K_{\beta \ell}
  Y_{\ell m}(\theta,\phi) \langle\beta \ell m\;|\;\beta  s\rangle X_{\beta s})
\end{equation}
The $a_{\ell'm'}$ coefficients are found from this by expanding
the temperature on spherical harmonics.
In the Appendix is shown that utilizing the expressions
(\ref{eq:m7}) for the ensemble averages of the random variables,
the ensemble average of $C_\ell$ may be expressed:
\begin{equation}  \label{eq:m15}
 \langle C_{\ell} \rangle
 = \sum_{m} \frac{1}{2 \ell+1} \langle a_{\ell m} \overline{a_{\ell m}} \rangle
 = \sum_{\beta} \frac{K_{\beta \ell}^2}{2} \frac{trace[G_{\beta} P_{\beta}(\ell)]}{2 \ell+1}
\end{equation}
where $G_{\beta}$ and $P_{\beta}(\ell)$ are projection operators
that projects the whole eigenspace to the Laplacian belonging to
the eigenvalue $\beta$ down to the eigenspace of group-symmetrical
functions, and to the subspace of a particular value of the
angular momentum $\ell$, respectively:
\begin{equation} \label{eq:m16}
G_{\beta} = \sum_{s} |\beta s \rangle \langle \beta s |
\qquad P_{\beta}(\ell) = \sum_{m}|\beta \ell m \rangle \langle \beta \ell m |
\end{equation}
Using that the operator $G_{\beta}$ is actually a sum of right
Clifford translations, and hence commutes with all left-screw
transformations \cite{gausmann}, as well as using the symmetries of the
$P_{\beta}(\ell)$ operator it is proven in the Appendix, that the
expression (\ref{eq:m15}) can be simplified to
\begin{eqnarray} \label{eq:m17}
  \langle C_\ell\rangle = \sum_{\beta > \ell} \frac{{|K_{\beta \ell}|}^2}{2}
  \;\frac{\mathrm{multiplicity}(\beta)}{\beta^2}
\end{eqnarray}
For $S^3$ we get just
\begin{eqnarray} \label{eq:m18}
  \langle C_\ell\rangle
  = \sum_{\beta > \ell} \frac{{|K_{\beta \ell}|}^2}{2}
\end{eqnarray}
The significance of (\ref{eq:m17}) is, that it shows that we do
not have to calculate anything but the radial function, and its
folding in equation (\ref{eq:m12}), to calculate the spectrum,
even for spaces with nontrivial topologies. The only way the
topology makes the result differ from the case of $S^3$ is through
the second factor in (\ref{eq:m17}). The $multiplicity(\beta)$ is
the dimension of the space of invariant eigenfunctions, respecting
the holonomies $\Gamma$ for each eigenvalue $-(\beta^2 -1)$ of the
Laplacian, and is known explicitly \cite{lehoucq}.
\\

For the variance of the $C_{\ell}$'s it is shown in the Appendix
that it is related to the matrix elements of the group
symmetrizing projection operator $G_{\beta}$ as follows:
\begin{equation}  \label{eq:m19}
Q_{\ell \ell'} =  \langle C_{\ell} C_{\ell'} \rangle -\langle C_{\ell}\rangle \langle C_{\ell'} \rangle
= \frac{1}{2}\frac{1}{2\ell+1}\frac{1}{2\ell'+1} \sum_{m m'} {|M_{m,m'}^{\ell \ell'}|}^2
\end{equation}
where the matrix $M$ is derived by "symmetrizing" the sum of the
matrix elements of the group averaging operator, as follows:
\begin{equation}  \label{eq:m20}
 \fl M_{m,m'}^{\ell \ell'}
  = \sum_{\beta} K_{\beta \ell}K_{\beta \ell'}
    \frac{\langle \beta \ell m|G_{\beta}|\beta \ell' m' \rangle
  + (-1)^{\ell+\ell'+m+m'}\langle \beta \ell -m|G_{\beta}|\beta \ell' -m' \rangle}{2}
\end{equation}
As shown in the Appendix, the matrix elements of $G_{\beta}$ can
be found solving a certain eigenvalue problem based on the fact
that the group symmetrical functions are invariant to the
holonomies of each manifold $S^3$/$\Gamma$. The matrix elements of
the hermitian operator $G_{\beta}$ are real, and hence symmetric.
As another characteristicum, we note that all the elements of the
covariance matrix $Q_{\ell \ell'}$ are non-negative. For $S^3$,
$G_{\beta}$ is diagonal, and we recover the familiar result:
\begin{equation}  \label{eq:m21}
 {Q^{S^3}}_{\ell \ell'}
 = \langle C_{\ell} C_{\ell'} \rangle -\langle C_{\ell}\rangle \langle C_{\ell'} \rangle
 = \frac{2}{2\ell+1}\delta_{\ell \ell'}{C_{\ell}}^2
\end{equation}
In practical terms, we proceed as follows, which is explained
further in the Appendix: For each grid point
$(\Omega_{tot},\Omega_{mass})$ in a 41 by 21 cell grid spanning
the range $\Omega_{tot}=1.004$ to $1.084$ and $\Omega_{mass}= 0.2$
to $0.4$, we solve the Friedmann equation (\ref{eq:friedmann})
from $x=1/1085$ (assuming the last scattering occurred at a
redshift of $z=1085$) and the equation (\ref{eq:m5}) for $F$ and
$F'$ numerically, which also supplies $\eta_{LSS}$, the conformal
time at the last scattering surface. Also, for each $\beta=3..101$
and $\ell=2...15$ we calculate the radial wavefunction in 400
points between 0 and $\pi$, using a symbolic calculator to expand
the Legendre functions. Interpolating $F$ and $F'$, we calculate
the integrand in the second factor of (\ref{eq:m12}) for the same
400 values of $\eta_0 -\eta$. Doing then the integral by
multiplication with a precalculated 400 by 400 matrix and
interpolating we get both the first and the second term in
(\ref{eq:m12}), for each cosmic grid point. From the $K_{\beta
\ell}$ values we then calculate the spectrum from (\ref{eq:m17})
and the variance from (\ref{eq:m19}) and (\ref{eq:m20}).
\\

From the spectrum and the variance we construct an approximate
likelihood function over the $(\Omega_{tot},\Omega_{mass})$
parameter space. The spectrum and the variance is further used to
calculate the two-point angular correlation function and its
variance, for each grid point.
\\

In the Appendices, the methods and mathematical relations
exploited in this work, are explained further, along with some
additional results.

\section{Results} \label{sec:results}

The best-fit WMAP models are a natural starting point, as they
incorporate a lot of constraints from the high $\ell$ multipoles
of the spectrum. As noted in the report on the first year results
from the WMAP team \cite{spergel} the position of the first
acoustic peak constrains the universe to be nearly flat. However,
for models with a nonzero cosmological constant there is a
geometric degeneracy along the lines of constant conformal
distance to the last scattering surface in the $\Omega_{mass}$
$\Omega_{tot}$ plane, which allows models with topology $S^3$ or
$S^3/\Gamma$ studied in this paper. Actually \cite{spergel} gives
as the best estimate for $\Omega_{tot}$ the value 1.02 +/- 0.02
which seems to favour a closed universe, but does not exclude
either a flat universe nor an open universe. The degeneracy means,
that the spectrum can be equally well fitted by assuming a flat as
well as a slightly curved universe.

It is therefore assumed in this paper that the modelled spectrum of
the WMAP-team coincides with the spectrum for $S^3$ for low
$\ell$'s, which is in fact verified. If the universe has a
non-trivial global topology, the high $\ell$ behaviour is
asymptotically the same if the topology is instead $S^3$/$\Gamma$,
provided a scaling of $|\Gamma|$ is applied to the spectrum of the
$S^3$/$\Gamma$ models.
\\

Indeed, scaling the spectrum for $S^3$ calculated by the methods
of this paper, to the WMAP modelled spectrum (we choose to use the
simple pl-model here), shows a very good consistency. Applying
next a relative scaling of $|\Gamma|$ to the spectrum of the
$S^3$/$\Gamma$ models studied here, we can check whether the
alternative topologies can provide a better fit to the low $\ell$
behaviour, while still keeping consistency with the WMAP models
high $\ell$ behaviour. As shown in Figure \ref{Figure2}, the $S^3$
model fits equally well over the different values of
$\Omega_{tot}$ so that the ex post likelihood distribution hardly
changes the ex ante prior used (a Gaussian centred at
$\Omega_{tot} = 1.02$ with width equal to the standard deviation
of $\Omega_{tot}$ of $0.02$ reported by the WMAP team). This
indicates that the procedure of scaling the overall amplitude of
the modelled $S^3$ spectrum to the WMAP model makes sense. For
$S^3/I^*$ and $S^3/O^*$, the ex post likelihood function becomes
much sharper than the prior distribution, enabling us potentially
to use the low multipoles to constrain the cosmological parameters
more than achieved by the WMAP team, whereas the potential
conclusions in the case of $S^3/T^*$ are more mixed. The much
sharper distribution over $\Omega_{tot}$ especially for $S^3/I^*$,
means that if the issue of the topology is settled, the
cosmological parameters might be more strictly constrained. For
$S^3/I^*$ the position and width of the two peaks in the
likelihood occurs at $\Omega_{tot}$ = 1.028 +/- 0.0023 and for the
smaller peak, at $\Omega_{tot}$ = 1.017 +/- 0.0015, almost the
same value found in \cite{aurich1} by studying the S-statistic
introduced in \cite{spergel} and explained in the \ref{sec:11} of
this paper. As this statistic is heavily biased to explain what we
see in the observations for the lowest multipole moments, we
favour instead the value 1.028 of the right peak, as the most
likely estimate.
\\

The conformal distance to the last scattering surface in the maximum
likelihood peak of the map for $S^3/I^*$ is found to be $\eta=0.571$
which implies that any matching circles in the sky would have a radius
of
\begin{equation}
\theta = acos \left( \sqrt{1-2 \sqrt{1/5}}\;\cot(\eta)\right) = {59.6}^{o}
\end{equation}
We {\it could} consider the topology to be a discrete cosmological
parameter, and find the relative ex post probabilities for each
one, by applying the Baysian principle, as we do for the
cosmological parameters. Assuming an apriori distribution of equal
ex ante probability for each topology, the ex post probability,
given the data, is very much in favour of the non-trivial
topologies. In fact we find by such a recipe the following ex post
probabilities:
\begin{equation}
\fl P(S^3/I^*)= 0.45 \; \;\; P(S^3/O^*)= 0.31\; \;\; P(S^3/T^*)= 0.13\; \;\; P(S^3)= 0.12
\end{equation}
This reasoning, however has several weaknesses. First of all, the
choice of using equal apriori probabilities is a highly subjective
choice. And secondly, models of inflation suggest an almost, if
not completely flat Universe. Nevertheless, it is
thought-provoking that the likelihood of the $S^3/I^*$ topology
has a sharp peak very near the values of
$\Omega_{mass},\Omega_{tot}$ found from the high $\ell$ data.
\\

In the case where we instead apply a uniform primer $P^{ante}$
over the $(\Omega_{tot},\Omega_{mass})$-window, we can not assume
that the modelled spectra should be scaled to the WMAP model.
Instead we have to resort to scaling the spectra to the observed
spectrum (which we know will produce bad results for $S^3$ as the
low-$\ell$ multipoles of the observations are systematically too
low, at least if a fit to higher $\ell$-multipoles shall be
realised as well). We are here bothered by the fact, that we have
not in this exercise, for computational reasons, modelled the
higher $\ell$ multipoles, which might more reliably make a fit
between the $S^3$ model and data meaningful (as all topologies
should approach the same curve asymptotically). Scaling
nevertheless mechanically to the observed spectrum, we arrive at
the results shown in Figure \ref{Figure3}.
\\

It is obvious, that the models $S^3/I^*$, $S^3/O^*$ and $S^3/T^*$
tend to have their optimum likelihood along the well-known
geometrical degeneracy line of constant distance (in conformal
time) to the last scattering surface. But except for $S^3/I^*$,
the maximum likelihood regions lye at relatively improbable values
of the cosmological parameters, i.e. at high $\Omega_{mass}$ or
$\Omega_{tot}$, and even outside the window in parameter space
studied here.
\\

\begin{figure}
\centerline{\includegraphics[width=15cm]{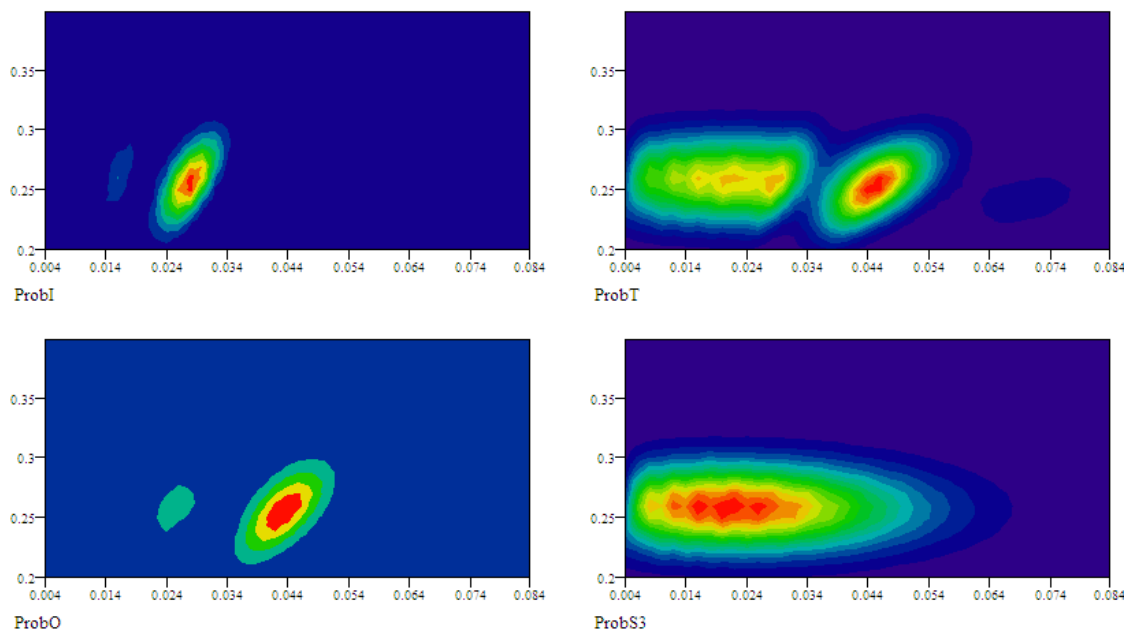}}
\caption{The ex post likelihood distribution over $\Omega_{mass}$
(vertical axis)  and $\Omega_{tot}-1$ (horizontal axis) for
$S^3/I^*$, $S^3/T^*$, $S^3/O^*$ and $S^3$, using a Gaussian
primer} \label{Figure1}
\end{figure}
\begin{figure}
\centerline{\includegraphics[width=15cm]{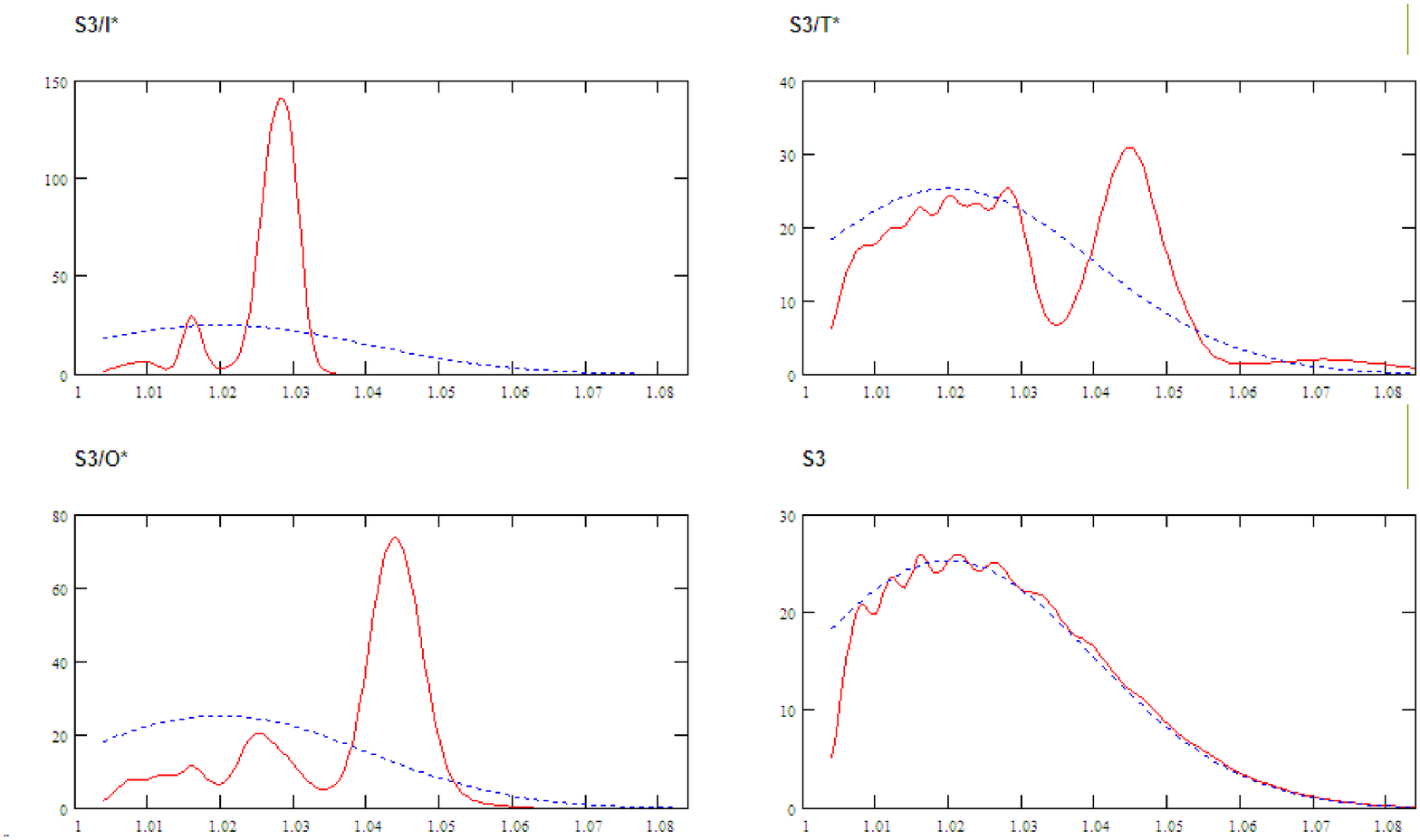}} 
\caption{The
likelihood distribution as a function of $\Omega_{tot}$ along the
most likely value of $\Omega_{mass}$ = 0.26 for $S^3/I^*$,
$S^3/O^*$, $S^3$ and = 0.25 for $S^3/T^*$. Dotted line shows the
prior distribution, for comparison. Note the different scales on
the vertical axes.} \label{Figure2}
\end{figure}

\begin{figure}
\centerline{\includegraphics[width=15cm]{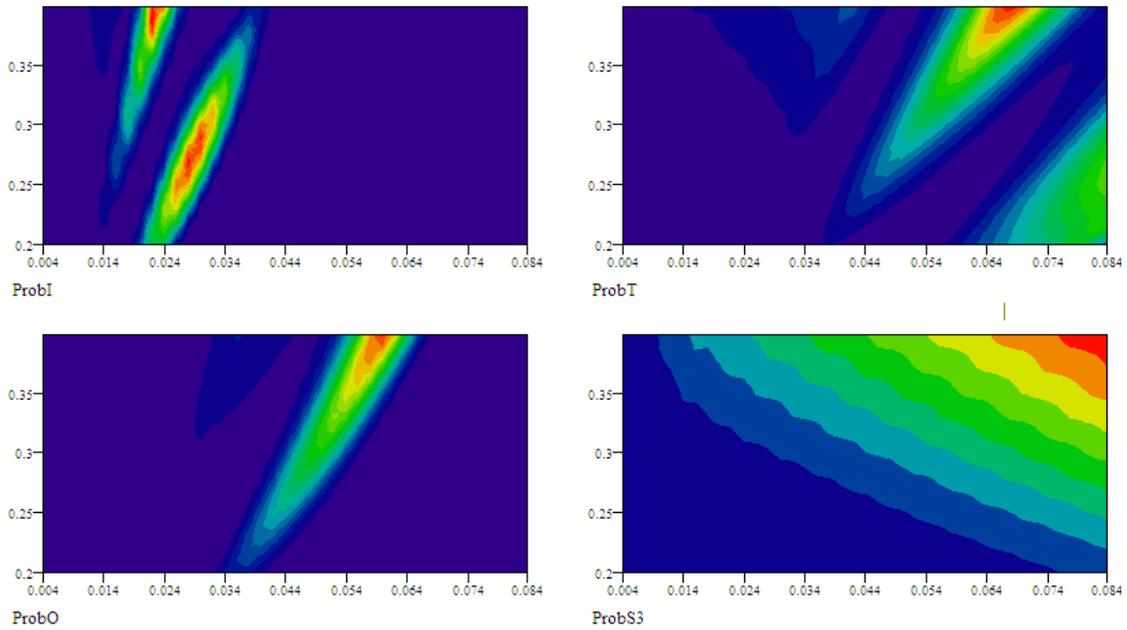}}
\caption{The ex post likelihood distribution over $\Omega_{mass}$
(vertical axis)  and $\Omega_{tot}-1$ (horizontal axis) for
$S^3/I^*$, $S^3/T^*$, $S^3/O^*$ and $S^3$, using a uniform primer,
and using best scaling to the observed multipole moments in the
range of $C_\ell$ = 2 to 15} \label{Figure3}
\end{figure}

\begin{figure}
\centerline{\includegraphics[width=15cm]{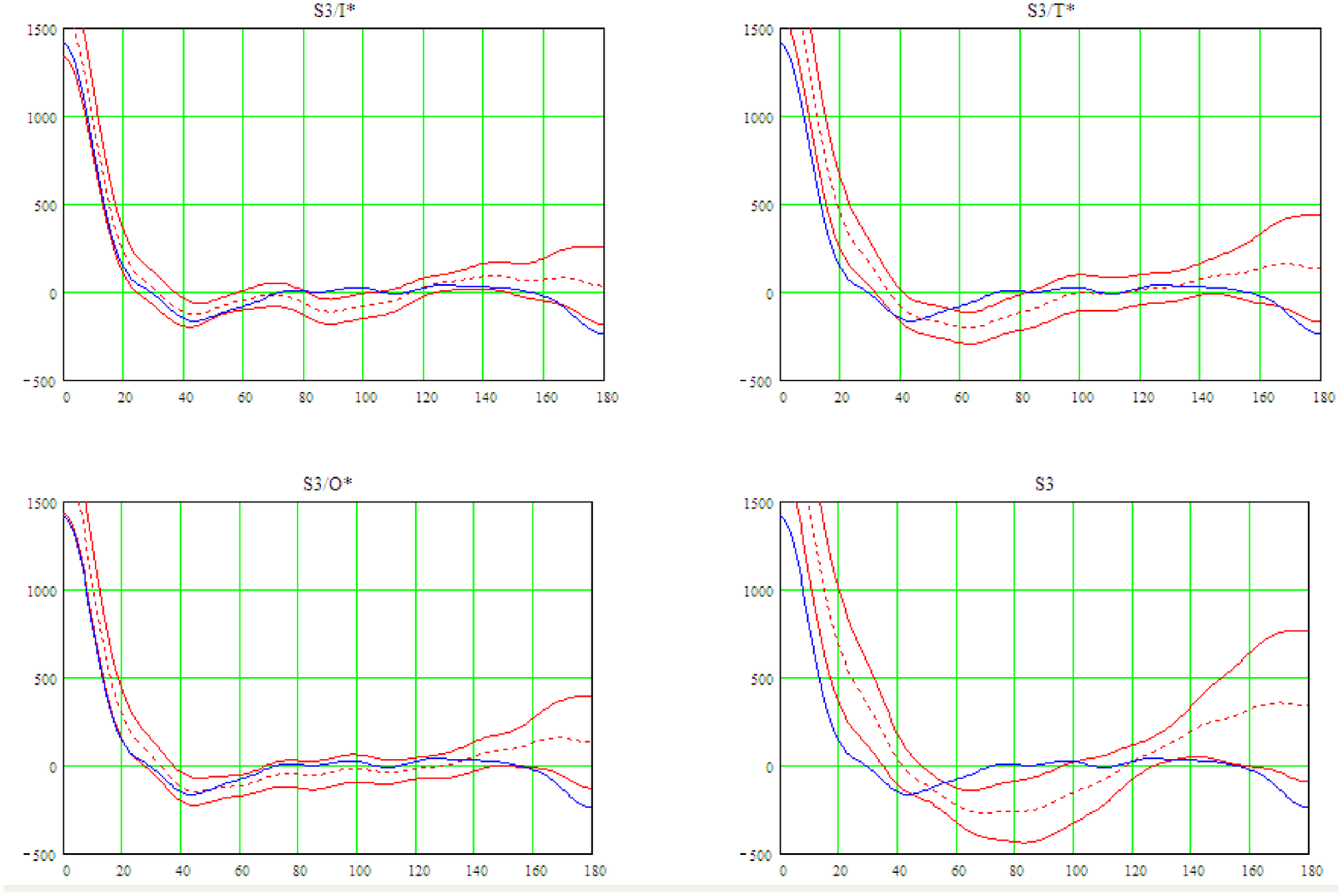}} \caption{The
temperature correlation function in the point of maximum
likelihood for each of the spaces (compare Figure \ref{Figure1}).
For $S^3/I^*$
 at $\Omega_{mass}=0.26$ and $\Omega_{tot}=1.028$, for $S^3/T^*$ at $\Omega_{mass}=0.25$ and $\Omega_{tot}=1.044$,
for $S^3/O^*$ at $\Omega_{mass}=0.26$ and $\Omega_{tot}=1.044$,
and for $S^3$ at $\Omega_{mass}=0.26$ and $\Omega_{tot}=1.016$.
The graph shows the modelled ensemble average with its one
standard deviation band, along with the observed
correlation-function, both filtered to a max $\ell$ of 15.}
\label{Figure4}
\end{figure}

\begin{figure}
\centerline{\includegraphics[width=15cm]{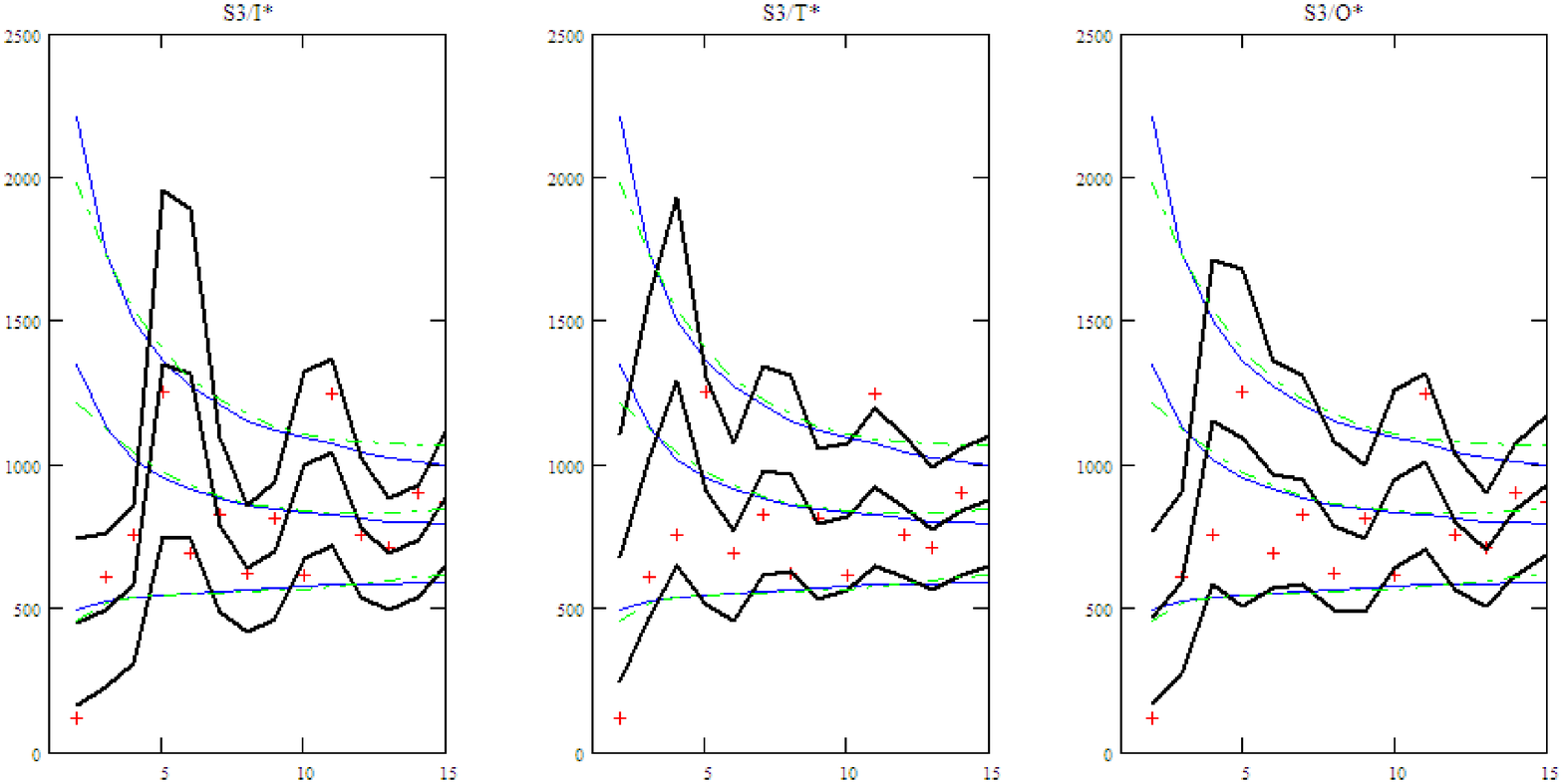}} \caption{The
spectrum for $S^3$/$I^*$, $S^3/T^*$ and $S^3/O^*$ for the optimum
likelihood values of the cosmological parameters (as in Figure
\ref{Figure4}). Fat curves show the spectrum +/- one standard
deviation (i.e. the square root of the diagonal elements of the
variance-matrix). Dotted line shows the WMAP model +/- the
calculated cosmological variance, and thin line shows the $S^3$
model +/- its standard deviation in its optimum likelihood point
of parameter space. Crosses show observed multipole moments}
\label{Figure5}
\end{figure}

The result of this procedure is shown in Figure \ref{Figure4} for
the case of the $S^3$/$I^*$ topology, for the case of the
so-called pl-model, which assumes a power law fit to the spectrum,
using the WMAP best estimates of the $\Omega_{mass}$ and
$\Omega_{tot}$ for this model.
\\

As seen, the angular correlation function is constrained almost to
within one standard deviation, except around 90 degrees and near
180 degrees, which is explained solely by the very low quadrupole
moment of the observations. Overall, however, the $S^3$/$I^*$
reproduces the observed correlation function much better than the
$S^3$ model.
\\

This is also shown in Figure \ref{Figure5}, showing the first 15
multipole moments, where the observed values are almost fairly
well constrained to the ensemble average of the model, +/- 1 times
the square root of the diagonal elements of the covariance matrix,
for the $S^3/I^*$ model, but is far outside 1 standard deviation
for the $S^3$ model, for the $\ell=2$ and $\ell=11$ moments.

\ack

The author is most grateful for helpful hints from key persons in
the international community working with multiconnected spaces and
the CMB, inter alia Roland Lehoucq, Marc Lachi\`eze-Rey, and Jeff
Weeks. Especially thanks to Jeff Weeks, for many thoughtful
comments and questions, which have helped to finalise this paper.
Any errors or shortcomings remain however the sole responsibility
of the author.


\appendix

\section*{Appendices}
\setcounter{section}{0}

\section{Rotations on $S^3$} \label{sec:1}

Rotations on $S^3$ can be described in complete analogy with usual
rotations in $E^3$, by considering the rotations to take place in
the 4-dimensional Euclidian embedding space for $S^3$. Working
with the variables $x_n$, $n=0,1,2,3$ $\bx=(x,y,z,w)$ a
counterclockwise infinitesimal rotation \emph{of the coordinate
system} in the xy-plane an angle $d\phi$ changes the coordinates
of $\bx$ into $\bx'$
\begin{equation}\label{eq:1}
\bx' = \left( \begin{array}{cccc}
1 & d\phi & 0 & 0 \\
-d\phi & 1 & 0 & 0 \\
0 & 0 & 1 & 0 \\
0 & 0 & 0 & 1 \end{array} \right)  \bx
\end{equation}
A scalar function $\psi(\bx)$ transforms hereby into
$\psi'(\bx')=\psi(\bx)$, as follows:
\begin{equation}\label{eq:2}
\psi'(\bx)=(1+i\,d\phi J_{01})\psi(\bx)
\end{equation}
where $J_{01}$ , the generator of the infinitesimal rotation, is
\begin{equation}   \label{eq:3}
J_{01}=-i (x \frac{d}{dy}-y \frac{d}{dx})
\end{equation}
Considering alternatively $\psi$ to be a function of the
coordinates $\alpha,\theta,\phi$,
\begin{equation}
\scriptstyle \left( \begin{array}{c}
\scriptstyle x  \\
\scriptstyle y  \\
\scriptstyle z  \\
\scriptstyle w  \end{array} \right) =
\scriptstyle \left( \begin{array}{c}
\scriptstyle sin(\alpha)sin(\theta)cos(\phi)  \\
\scriptstyle sin(\alpha)sin(\theta)sin(\phi) \\
\scriptstyle sin(\alpha)cos(\theta)  \\
\scriptstyle cos(\alpha)  \end{array} \right)
\end{equation}
one finds that
\begin{equation}\label{eq:4}
J_{01}=-i  \frac{d}{d\phi}.
\end{equation}
A finite rotation \emph{of the coordinate system} through the
angle $\phi$ transforms $\psi$ as follows
\begin{equation}\label{eq:5}
\psi'(\bx)=e^{i \phi J_{01}}\psi(\bx).
\end{equation}
If instead we think of examining the value of a function at a
position $\bx'$ that arises by \emph{rotating the vector} $\bx$
the angle $\phi$, ie
\begin{equation}\label{eq:6}
\bx ' = R_{xy}(\phi) \bx \qquad
R_{xy}(\phi) = \left( \begin{array}{cccc}
cos(\phi) & -sin(\phi) & 0 & 0 \\
sin(\phi) & cos(\phi) & 0 & 0 \\
0 & 0 & 1 & 0 \\
0 & 0 & 0 & 1 \end{array} \right)
\end{equation}
we can do it by rotating the coordinate system the same amount,
finding the transformed function, and checking its value at the
new coordinate equal to the old, i.e.
\begin{equation} \label{eq:7}
\psi(R_{xy}(\phi)\cdot \bx)=e^{i \phi J_{01}}\psi(\bx)
\end{equation}
If we do two consecutive rotations we have to undo them in reverse
order:
\begin{equation} \label{eq:8}
\psi(R_{zx}(\theta)R_{xy}(\phi) \cdot \bx) = e^{i\theta J_{21}}\psi(R_{xy}(\phi) \cdot \bx)
= e^{i\phi J_{01}}e^{i\theta J_{21}}\psi(\bx)
\end{equation}

\section{The Lie algebra of rotations on $S^3$}  \label{sec:2}

In the following we state some useful properties of the Lie
algebra of rotations on $S^3$. The generators $J_{k\ell} =
-J_{\ell k}\;, \ell \ne k = 0, \ldots ,3$ of the rotations are
hermitian and satisfy the following commutation relations:
\begin{equation}\label{eq:9}
[J_{k\ell},J_{\ell m}] = i J_{m k}
\end{equation}
for any three different indices $k,\ell,m$ as well as
\begin{equation}\label{eq:10}
[J_{k\ell},J_{m n}] = 0
\end{equation}
when $k,\ell,m$ and $n$ are all different.
\\

For an observer sitting in $(x,y,z,w)=(0,0,0,1)$ the three
generators $J_{01}$, $J_{20}$ and $J_{12}$ will be the analogue on
$S^3$ for usual rotations around the z-, y and x-axis
respectively, and the sum of their squares will be the analogue of
the familiar total angular momentum $L^2$.
\begin{equation}\label{eq:11}
L^2 = J_{01}^2+J_{20}^2+J_{12}^2
\end{equation}
It is useful to introduce also the right and left screw
generators:
\begin{equation}\label{eq:12}
SR_0=J_{12}+J_{30} \qquad SR_1=J_{20}+J_{31}\qquad SR_2=J_{01}+J_{32}
\end{equation}
\begin{equation}\label{eq:13}
SL_0=J_{12}-J_{30} \qquad SL_1=J_{20}-J_{31}\qquad SL_2=J_{01}-J_{32}
\end{equation}
We note, that the left screw generators here are chosen with the
opposite sign than if thought of as mowing from
$(x,y,z,w)=(0,0,0,1)$ in the direction of the z-axis and
simultaneously making a left-hand rotation of y towards x. This
makes the following formulas more symmetric between the $SR$ and
$SL$ set.
\\

The screw generators satisfy the following commutation relations:
\begin{equation}\label{eq:14}
[SR_0,SR_1]= 2 i SR_2 \qquad  [SL_0,SL_1]= 2 i SL_2
\end{equation}
and analogous, by cyclic permutation.
\\

Further, any pair of left and right screw generators commute:
\begin{equation} \label{eq:15}
[SL_k,SR_\ell]= 0
\end{equation}
The following operators commute with all the generators, and are
thus Casimir-operators of the Lie group:
\begin{equation}\label{eq:16}
J^2 = J_{01}^{\phantom{01}2}
    + J_{02}^{\phantom{02}2}+J_{03}^{\phantom{03}2}
    + J_{12}^{\phantom{12}2}+J_{13}^{\phantom{13}2}
    + J_{23}^{\phantom{23}2}
  = \frac{1}{2} ( SL^2 + SR^2 )
\end{equation}
\begin{equation}\label{eq:17}
SR^2 = SR_{0}^{\phantom{0}2}+SR_{1}^{\phantom{1}2}+SR_{2}^{\phantom{2}2}
\end{equation}
\begin{equation}\label{eq:18}
SL^2 = SL_{0}^{\phantom{0}2}+SL_{1}^{\phantom{1}2}+SL_{2}^{\phantom{2}2}
\end{equation}
Further, we can construct lifting operators
\begin{equation}\label{eq:19}
AR_2 = SR_{0}+ i SR_{1} \qquad SR_2 \cdot AR_2 = AR_2 \cdot SR_2 + 2 \cdot AR_2
\end{equation}
\begin{equation}\label{eq:20}
AL_2 = SL_{0}+ i SL_{1} \qquad SL_2 \cdot AL_2 = AL_2 \cdot SL_2 + 2 \cdot AL_2
\end{equation}
These relations (that holds also by cyclic permutation) show that
$AR_2$ acting on an eigenstate to $SR_2$ with eigenvalue $sr$ is
also an eigenstate to $SR_2$ with eigenvalue $sr+2$, i.e. $AR_2$
"lifts" the eigenvalue of $SR_2$ 2 units, and analogously for
$AL_2$ acting on an eigenstate to $SL_2$. In a similar way, the
adjoint operators are seen to lower the eigenvalue 2 units:
\begin{equation}\label{eq:21}
AR_2^\dag = SR_{0}- i SR_{1} \qquad SR_2 \cdot AR_2^\dag = AR_2^\dag \cdot SR_2 - 2 \cdot AR_2^\dag
\end{equation}
and analogously for $AL_2^\dag$.
\\

The left screw lifting and lowering operator does not change the
eigenvalue of the right screw operator and vice versa:
\begin{equation}\label{eq:22}
[AL_2,SR_2] = [AL_2^\dag,SR_2] = [AR_2,SL_2] = [AR_2^\dag,SL_2] = 0
\end{equation}
We can thus have states that are simultaneously eigenvectors to
$SL_2$ and $SR_2$, with eigenvalues in a range
$sl_{min},sl_{min}+2 \ldots sl_{max}$  determined by the
nonvanishing of the result of the lifting and lowering operations.
The squared norm of a state resulting from application of the
lifting operator is (we here use the Dirac-notation explained in
\ref{sec:3}):
\begin{equation}\label{eq:23}
\langle sl\; sr |AL_2^\dag AL_2|sl\; sr \rangle =
\langle sl\; sr |SL_0^{\phantom{0}2}+SL_1^{\phantom{1}2}+i[SL_0,SL_1]|sl\; sr \rangle= \nonumber
\end{equation}
\begin{equation}\label{eq:24}
\langle sl\; sr |SL^2-SL_2^{\phantom{2}2}-2 SL_2|sl\; sr \rangle = SL^2-sl^2-2 sl
\end{equation}
where we have used that $SL^2$ commutes with all the generators and thus is a common eigenvalue
for all the states.
\\

Similarly, we find that the squared norm after lowering is
\begin{equation}\label{eq:25}
\langle sl\; sr |AL_2 AL_2^\dag|sl\; sr \rangle = SL^2-sl^2+ 2 sl
\end{equation}
We thus see that the number $\beta_l$ of different eigenvalues
$sl=-(\beta_l-1),-(\beta_l-1)+2, \ldots,\beta_l-1$ is related to
$SL^2$ as follows $SL^2=\beta_l^{\phantom{l}2}-1$. We would get
the same result by examining the possible range of eigenvalues for
any of the left screw operators, so the multiplet of states should
also be spanned by a basis of eigenstates of the other two left
screw generators as well, with $\beta_l$ different eigenvalues. A
similar reasoning shows that the range of possible eigenvalues for
$SR_2$ is $sr=-(\beta_r-1),-(\beta_r-1)+2, \ldots,\beta_r-1$ where
$\beta_r$ is the number of different eigenvalues, related to
$SR^2$ as follows $SR^2=\beta_r^{\phantom{r}2}-1$.
\\

The 2 numbers $\beta_l$ and $\beta_r$ completely characterise the
multiplet, and we can easily construct a matrix representation of
the operators, by choosing a basis where $SL_2$ and $SR_2$ are
diagonal. The dimension of the representation is equal to $\beta_l
\cdot \beta_r$. In such a representation the nonvanishing matrix
elements can be chosen as
\begin{equation}\label{eq:26}
\langle sl\;sr|SL_2|sl\;sr \rangle = sl \qquad \langle sl\;sr|SR_2|sl\;sr\rangle = sr \nonumber
\end{equation}
\begin{equation}\label{eq:27}
\langle sl+2\;sr|AL_2|sl\;sr \rangle = i \sqrt{\beta_l^{\phantom{l}2}-sl^2-2 sl}=al(sl) \nonumber
\end{equation}
\begin{equation}\label{eq:28}
 \langle sl\;sr+2|AR_2|sl\;sr \rangle = i \sqrt{\beta_r^{\phantom{r}2}-sr^2-2 sr}=ar(sr)
\end{equation}
The above analysis illustrates that any irreducible representation
of the commutator-algebra of the generators $J_{k \ell}$ of
rotations of $E^4$ is in fact the direct product of
representations of the algebras of generators for rotating $E^3$:
$SO(4) = SO(3)\times SO(3)$, which is further exploited in
\ref{sec:5} (see also \cite{kramer} which uses this fact to derive
explicit analytical solutions for the lowest eigenfunctions of the
Poincar\'e dodecahedral space). Although we can have abstract
multiplets for any $\beta_r,\beta_l$, the multiplets corresponding
to single component wave functions have $\beta_r = \beta_l =
\beta$. One may in fact show, that the Laplacian for such wave
functions is simply $-J^2$. There are thus $\beta^2$ states in the
multiplet, corresponding to the multiplicity $\beta^2$ for
eigenstates to the Laplacian on $S^3$, with eigenvalue
$-(\beta^2-1)$. The usefulness of studying the possible choices of
commuting operators is that we can use the eigenvalues of the
operators to label the degenerate eigenstates of the Laplacian on
$S^3$. Where the conventional choice is to use the set of
operators $J^2$,$L^2$,$J_{01}$, we have found it useful to employ
instead the set $J^2$, $SR_2$ and $SL_2$.
\\

The above explicit expressions (\ref{eq:26}) to (\ref{eq:28})
determine all the generators $J_{k\ell}$. There is a choice of
relative phases between states involved, in the choice of the
lifting operators. This is purely conventional, and the
expressions may be multiplied by any complex number with modulus
1. Our use of the factors of $i$ are convenient, however, as then
the operators $J_{01}$ and $J_{20}$ which are generators of
rotations around the z-axis and y-axis respectively are both real
matrices:
\begin{equation}\label{eq:29}
 SR_0 = \frac{1}{2} (AR_2+AR_2^\dag) \quad SL_0
      = \frac{1}{2} (AL_2+AL_2^\dag)
\end{equation}
\begin{equation}\label{eq:30}
 SR_1 = \frac{1}{2i} (AR_2-AR_2^\dag) \quad SL_1
      = \frac{1}{2i} (AL_2-AL_2^\dag)
\end{equation}
\begin{equation}\label{eq:31}
 J_{12} = \frac{1}{2} (SR_0+SL_0) \quad J_{20}
        = \frac{1}{2} (SR_1+SL_1) \quad J_{01} = \frac{1}{2} (SR_2+SL_2)
\end{equation}
\begin{equation}\label{eq:32}
 J_{30} = \frac{1}{2} (SR_0-SL_0) \quad J_{31}
        = \frac{1}{2} (SR_1-SL_1) \quad J_{32} = \frac{1}{2} (SR_2-SL_2).
\end{equation}

\section{The eigenstates of the Laplacian on $S^3$}  \label{sec:3}

We can in fact use the rotation operators to calculate the
eigenfunctions.  We use the Dirac notation \cite{dirac}, well
known from its use in quantum mechanics:
\begin{equation}\label{eq:33}
\langle \;f\;|\;g\;\rangle = \int
\overline{f(\alpha,\theta,\phi)}g(\alpha,\theta,\phi) d\Omega
\end{equation}
where $d\Omega=\sin(\alpha)^2 \sin(\theta)d\alpha d\theta d\phi$.
This expression is what physicists would call the expression for
$\langle f | g \rangle$ in the coordinate representation. If
however we have an expansion of $f$ and $g$ on a complete set of
basis functions, call them $|\beta\;\ell\;m\rangle$, we can also
calculate $\langle f | g \rangle$ in the following matrix
representation:
\begin{equation}\label{eq:34}
F(\beta,\ell,m)=\langle \beta \; \ell \; m | f \rangle \qquad G(
\beta , \ell , m)= \langle \beta \; \ell \; m | g \rangle
\end{equation}
as
\begin{equation}\label{eq:35}
\langle \;f\;|\;g\;\rangle=\sum_{ \beta \; \ell \; m} \overline{F(
\beta , \ell , m)} \; G( \beta , \ell , m)
\end{equation}
or for short
\begin{equation}\label{eq:36}
\langle \;f\;|\;g\;\rangle=F^\dag G
\end{equation}
Considering $|\;\alpha\;\theta\;\phi\;\rangle$ to be the name of
the delta-function on $S^3$, we can write a scalar function $\psi$
as
\begin{equation}\label{eq:37}
\psi(\alpha,\theta,\phi)= \langle\; \alpha\;\theta\;\phi\;|\;\psi\;\rangle
\end{equation}
An eigenfunction to the Laplacian, say the $\beta,\ell,m$ state,
may then be calculated using the rotation operators, as follows
\begin{equation}\label{eq:38}
\psi_{\beta\ell m}(\alpha,\theta,\phi)= \langle \; \alpha \;
\theta \; \phi \;|\;\beta\;\ell\;m\;\rangle = \langle
\;0\;\;0\;\;0\;|\;e^{i \alpha J_{32}}e^{i \theta J_{20}}e^{i \phi
J_{01}}\;|\;\beta\;\ell\;m\;\rangle
\end{equation}
This expression may most easily be understood, by noting that we
simply make rotations of the coordinate system in the order
$\phi,\theta,\alpha$ that has the effect that the point
$(\alpha,\theta,\phi)$ gets new coordinates $(0,0,0)$
\begin{equation}\label{eq:39}
\psi'=e^{i \alpha J_{32}}e^{i \theta J_{20}}e^{i \phi J_{01}}\psi
\qquad \psi'(0,0,0)=\psi(\alpha,\theta,\phi)
\end{equation}
which is maybe easier to grasp than the alternative
\begin{equation}\label{eq:40}
\langle\;0\;0\;0\;|\;e^{i \alpha J_{32}}e^{i \theta J_{20}}e^{i \phi J_{01}}
=\langle\;\alpha\;\theta\;\phi\;|
\end{equation}
which in conjugated form is
\begin{equation}\label{eq:41}
|\;\alpha\;\theta\;\phi\;\rangle = e^{-i \phi J_{01}}e^{- i \theta
J_{20}}e^{- i \alpha J_{32}}|\;0\;0\;0\;\rangle
\end{equation}
and which can be seen as the $|\;\alpha\;\theta\;\phi\;\rangle$
state arising from a rotation of the state $|\;0\;0\;0\;\rangle$
in the sequence $\alpha,\theta,\phi$, where there is a minus as we
rotate the state, rather than the coordinate system.
\\

How do we calculate the expression (\ref{eq:38})? Well, the rule,
which is so familiar in quantum mechanics, is that the only way to
calculate the result of a linear operator, defined as a function
of another linear operator, is to do it in a basis of
eigenfunctions. So each of the above rotations must be calculated
in a basis of eigenstates for the $J_{32},J_{20}$ and $J_{01}$
operators, respectively, making it necessary to transform between
these bases.
\\

These rotations mix the various eigenstates belonging to a
beta-subspace, but do not mix states with different betas. In the
calculation of the eigenfunctions we can therefore freely insert
projection operators such as $\sum_{sl sr}{|  sl sr\rangle\langle
sl sr |}$ or $\sum_{\ell m}{| \ell m\rangle\langle  \ell m|}$ for
that subspace. In that way we get from equation (\ref{eq:38}) (and
now suppressing the $\beta$-index everywhere)
\begin{eqnarray}\label{eq:42}
\fl \psi_{\beta\ell m}(\alpha,\theta,\phi)
=
 \sum_{\ell'', m'', sl, sr, sr',sl',\ell', m'}
 \langle 000| \beta\; \ell'' \; m''\;\rangle
 \langle \; \ell'' \; m''\; | sl sr \rangle
 \langle sl sr | e^{i \alpha J_{32}} | sl' sr' \rangle \nonumber\\
 \times \langle sl' sr' | \ell' m' \rangle
 \langle \ell' m' | e^{i \theta J_{20}}  e^{i \phi J_{01}} \; |\; \ell \; m \; \rangle
\end{eqnarray}
Simplifying, we get
\begin{eqnarray}\label{eq:43}
\psi_{\beta\ell m}(\alpha,\theta,\phi) =
 \sum_{ sr}
\langle 000 | \beta 00 \rangle \langle \ell''=0 \;m''=0|sl=-sr \;sr \rangle
e^{i \alpha \frac{sr-sl}{2}} \nonumber\\
\times \langle sl=-sr \;sr |\ell \; m'=0 \rangle \langle \ell \;
m'=0 | e^{i \theta J_{20}} e^{i \phi J_{01}} \; |\; \ell \; m \;
\rangle
\end{eqnarray}
We have used the diagonal character of the rotation operators,
acting in their respective bases, to set indices equal, where
appropriate. We have also used, that the only $\beta,\ell,m$
combination that gives a $\psi_{\beta\ell m}(0,0,0)\ne 0$ is
$\beta,\ell,m = \beta,0,0$. Further, the value is
\begin{equation}\label{eq:44}
\psi_{\beta 0 0}(0,0,0) = \langle 000 | \beta 00 \rangle =
i \; \sqrt{\frac{\beta^2}{2\pi^2}}
\end{equation}
We do not calculate (\ref{eq:38}) in the coordinate
representation, but instead set up a matrix representation, where
each state is a vector, and the operators are matrices.
\\

As a final twist, we can write (\ref{eq:43}) as
\begin{eqnarray}\label{eq:45}
& &\psi_{\beta\ell m}(\alpha,\theta,\phi) \nonumber\\
&=& \sum_{ sr}
 N_{\beta\ell} \langle \ell=0
 \;m=0|sl=-sr \;sr \rangle e^{i \alpha sr}
\langle sl=-sr \;sr | \ell \;m'=0\rangle  Y_{\ell m}(\theta,\phi) \nonumber\\
&=& R_{\beta \ell}(\alpha) Y_{\ell m}(\theta,\phi)
\end{eqnarray}
where
\begin{equation} \label{eq:45a}
N_{\beta\ell}= i \;\sqrt{\frac{2 \beta^2}{\pi(2\ell+1)}}
\end{equation}
It is seen, that we have derived a Fourier expansion of the radial
function $R_{\beta \ell}(\alpha)$. We here used the result for
$S^2$, analogous to (\ref{eq:38}), that
\begin{equation}
Y_{\ell m}(\theta,\phi)= \langle 00 | e^{i \theta J_{20}} e^{i \phi J_{01}} | \ell m \rangle
= \langle 00 | \ell 0 \rangle  \langle \ell 0 | e^{i \theta J_{20}} e^{i \phi J_{01}} | \ell m \rangle
\end{equation}
where
\begin{equation}
\langle 00|\ell 0 \rangle = \sqrt{\frac{2 \ell +1}{4 \pi}}
\end{equation}

\section{The group symmetrical states}  \label{sec:4}

For the manifolds $S^3/\Gamma$ we are studying, there exist a set
of Clifford translations, that connects any point to each of its
$|\Gamma|$ ghost-images. These Clifford translations are defined
as the screw-transformations that bring the origin to the images
of the origin. The Universe may employ either left or right screw translations,
which however is immaterial for the analysis of this paper.
Here and in the following, we settle
for right-screw transformations. It is convenient to imagine the
observer to be situated in the "origin" $(x,y,z,w)=(0,0,0,1)$, which has
angular coordinates $(\alpha,\theta,\phi)=(0,0,0)$.
\\

The coordinate transformation involved for the $\bx_k$ $k=0,1,2,3$
is thus, for a screw-translation $g$ which brings the origin to a
point with angular coordinates $(\alpha,\theta,\phi)$
\begin{equation}\label{eq:46}
\bx' = g(\bx)= R_{xy}(\phi) R_{zx}(\theta) R_{screw}(\alpha)
R_{zx}(-\theta) R_{xy}(-\phi) \; \bx
\end{equation}
Here, the rotation matrices $R_{xy}$ and $R_{zx}$ are the usual
$4 \times 4$ matrices for rotating a vector an angle $\phi$ and
$\theta$ in the xy plane and the zx plane respectively, leaving
the other two coordinates unchanged, while $R_{screw}$ performs a
right-screw transformation along the z-axis:
\begin{equation}\label{eq:47}
R_{screw}(\alpha) = \left( \begin{array}{cccc}
cos(\alpha) & -sin(\alpha) & 0 & 0 \\
sin(\alpha) & cos(\alpha) & 0 & 0 \\
0 & 0 & cos(\alpha) & sin(\alpha) \\
0 & 0 & -sin(\alpha) & cos(\alpha) \end{array} \right)
\end{equation}
The logic is, that
first we apply a rotation of the coordinate system (first $\phi$,
then $\theta$) that brings our target point to lie on the z-axis,
then we do the screw, and transforms back to the proper
$\theta,\phi$ direction.
\\

Instead of transforming the coordinates, we use our insights from
\ref{sec:1}, equations (\ref{eq:7}) and (\ref{eq:8}), that we can
alternatively find the transform of the function $\psi$
\begin{equation}\label{eq:48}
\psi(\bx')=\langle \bx'|\psi\rangle = \psi'(\bx)=\langle \bx|e^{-i
\phi J_{01}}e^{-i \theta J_{20}}e^{i \alpha SR_2}e^{i \theta
J_{20}}e^{i \phi J_{01}}|\psi\rangle
\end{equation}
We can make the average over all the $|\Gamma|$ ghost images of
\bx, and find that it is
\begin{equation}\label{eq:49}
\frac{1}{|\Gamma|}\sum_{g}\psi(g(\bx))=\langle
\bx|\frac{1}{|\Gamma|}\sum_{n} e^{-i \phi_n J_{01}}e^{-i \theta_n
J_{20}}e^{i \alpha_n SR_2}e^{i \theta_n J_{20}}e^{i \phi_n
J_{01}}|\psi\rangle
\end{equation}
where the sum is over all the coordinates $\alpha,\theta,\phi$ of
the ghost images of the origin. We see from this, that any
group-symmetrical function $\psi$, for which
$\psi(g(\bx))=\psi(\bx)$ for all $g$, evidently is an
eigenfunction to the group-averaging operator, with eigenvalue 1:
\begin{equation}\label{eq:50}
\langle \bx|\psi\rangle = \langle \bx|G_\beta|\psi \rangle
\end{equation}
where
\begin{equation} \label{eq:51}
G_\beta = \frac{1}{|\Gamma|}\sum_{n} e^{-i \phi_n J_{01}}e^{-i
\theta_n J_{20}}e^{i \alpha_n SR_2}e^{i \theta_n J_{20}}e^{i \phi_n
J_{01}}
\end{equation}
It may be shown, that the group averaging operator is in fact a
projection operator, i.e. an operator with eigenvalues either 0 or
1 \cite{lehoucq}. Finding the group symmetrical functions thus
boils down to finding the eigenvectors to the matrix $G_\beta$,
that have eigenvalue 1. Denoting these $|\beta\;s\rangle$ where
$s=1,\ldots,multiplicity(\beta)$, the permissible eigenfunctions
for the Laplacian on $S^3/\Gamma$ are
\begin{equation}\label{eq:52}
\psi_s(\bx)=\langle \bx | \beta\;s\rangle
\end{equation}
The fact that $G_\beta$ is a projection operator means that we can
write
\begin{equation}\label{eq:53}
G_\beta=\sum_{s}\;|\beta\;s\rangle \langle \beta\;s\;|
\end{equation}
The operator G has many symmetries. First of all, it is evident
from equation (\ref{eq:51}), that G is a sum of right-screw
Clifford-translations. As any right-screw Clifford translation
commutes with any left-screw Clifford translation \cite{gausmann} this has
the consequence, that G commutes with $SL_2$ as well as the
lifting-operator $AL_2$. This implies that eigenstates to G may be
chosen as simultaneous eigenstates to $SL_2$. Each such eigenstate
will then be a superposition of states with identical left-screw
eigenvalue but different right-screw eigenvalues:
\begin{equation}\label{eq:54}
|\beta s \rangle = |\beta sl s' \rangle = \sum_{sr} a_{s'}^{sr}| \beta sr sl \rangle
\end{equation}
Here we can choose $a_{s'}^{sr}$ to be independent of $sl$ as is
easily seen by acting with the lifting operator $AL_2$ on the sum.
This is seen to be consistent with the known fact that the
multiplicity, i.e. the number of eigenstates for each $\beta$, is
a multiple of $\beta$. Now the group-symmetrical functions must be
eigenstates with eigenvalue 1 to each of the Clifford-translations
in the group $\Gamma$ defining the space $S^3/\Gamma$.
Specifically, if we choose our coordinate system to have its
z-axis aligned along one of the basic Clifford-translations, they
must be eigenfunctions with eigenvalue 1 to the screw-translations
along the z-axis over an angle $\alpha_{\Gamma} =
\frac{2\pi}{N_{\Gamma}} = \frac{2\pi}{10}, \frac{2\pi}{8},
\frac{2\pi}{6}$ for the case of $\Gamma = I^*,O^*,T^*$
respectively.
\\

As each of the $sr,sl$ eigenstates is an eigenstate to such
translations with eigenvalue $e^{i \; sr \; \alpha_{\Gamma}}$ we
realise that the only admissible $sr$ values in the expansion
(\ref{eq:54}) are $sr=0$ mod $N_{\Gamma}$.

\section{The $\ell m$ states} \label{sec:5}

We can find the simultaneous eigenvectors $|\beta \;\ell \;m
\rangle$ of the operators $L^2$ and $J_{01}$ (the conventional
$\ell,m$ set) by simply calculating the eigenvectors of the matrix
$L^2+J_{01}$ the eigenvalues of which are all different. Our
software (Mathcad) then automatically supplies real eigenvectors,
as the matrix is real, meaning that the transformation matrix from
the $|sr,sl\rangle$ basis to the $|\beta\; \ell\; m\rangle$ basis
becomes real. We can fix the sign of each $\ell,m $ eigenvector,
by calculating the wave function (\ref{eq:38}) in a single point,
and comparing with the following analytical expression:
\begin{equation}\label{eq:55}
 \langle \alpha \theta \phi | \beta \;\ell \;m \rangle
  = i^{m}R_{\beta \ell}(\alpha) Y_{\ell m}(\theta,\phi)
\end{equation}
The factor of $i^m$ will disappear if we choose to omit the $i$ in
(27) and (28), which is the convention used in the following.
\\

Here $Y_{\ell m}$ are the usual spherical harmonic functions, with
the symmetry
\begin{equation}\label{eq:56}
\overline{Y_{\ell m}} = (-1)^m Y_{\ell -m}
\end{equation}
whereas the radial function $R_{\beta \ell}$ has the explicit
analytical expression \cite{riazuelo}:
\begin{equation}\label{eq:57}
 R_{\beta \ell}(\alpha)
  = M_{\beta \ell}
    \frac{PL(\beta-\frac{1}{2},-(\ell+\frac{1}{2}),cos(\alpha))}
         {(1-cos(\alpha)^2)^{\frac{1}{4}}}
\end{equation}
where the normalisation factor is
\begin{equation}\label{eq:58}
M_{\beta \ell} = i^{-\ell} \sqrt{\prod_{n=0}^{\ell} (\beta^2-n^2)}
\end{equation}
and $PL$ is the Legendre function
\begin{equation}\label{eq:59}
PL(n,m,u)=(1-u^2)^{\frac{m}{2}} \frac{(-1)^n}{2^n \Gamma(n+1)} \frac{d^{m+n}}{du^{m+n}} (1-u^2)^n
\end{equation}
Due to the factor of $(-1)^n$ with half-integer $n$, the Legendre
function is purely imaginary. Combined with the factor of
$i^{-\ell}$ of $M_{\beta \ell}$ the complex conjugate of our
radial function is
\begin{equation}\label{eq:60}
\overline{R_{\beta \ell}} = -(-1)^{\ell}R_{\beta \ell}
\end{equation}
This means that also
\begin{equation}\label{eq:60a}
\overline{K_{\beta \ell}} = -(-1)^{\ell}K_{\beta \ell}
\end{equation}
We note that an extra factor of $i^{-\ell}$ is used in this paper
compared to \cite{riazuelo} to get consistency with (\ref{eq:45}).

Using numerical eigenvector-determination to find the
transformation matrices $\langle sr \; sl|\ell \; m \rangle$ has
its limits, due to the huge size of the matrices ($\beta^2$ by
$\beta^2$ matrices).  Memory constraints thus limit the
feasibility to a max $\beta$ of 43, on a PC with 512 Mb ram. The
$\ell,m$ states can however easily be expanded analytically on the
$sr,sl$ states. To see this, note that the commutation relations
(14) show that considered as 3-vectors both $JL=\frac{SL}{2}$ and
$JR=\frac{SR}{2}$ satisfy the usual commutation relations for
angular momentum operators
\begin{equation}\label{eq:61}
[J_x,J_y]=i J_z
\end{equation}
Eigenstates of $SR_2$ and $SL_2$ are thus states with definite
values of the z-component of $JR$ and $JL$ as well as of their
absolute magnitude squared, $JR^2=JL^2=\frac{k}{2}(\frac{k}{2}+1)$
where $k=\beta-1$.
\\

As the components of the $JR$ and $JL$ commute (equation
(\ref{eq:15}) we can, from simultaneous eigenstates of $JR^2,JR_2$
and $JL^2,JL_2$ construct eigenstates of their sum,
$L=JR+JL=(J_{12},J_{20},J_{01})$, i.e. eigenstates of definite
$L^2=\ell(\ell+1)$ and $L_2=m$, by the rule for vector addition of
angular momentum \cite{landau}:
\begin{equation}\label{eq:62}
 | \ell m \rangle
 = \sum_{sr sl} (\ell, \frac{k}{2},\frac{k}{2};m,\frac{sr}{2},\frac{sl}{2})|sr sl \rangle
\end{equation}
This shows that the expansion coefficients between the eigenstates
$|sr,sl\rangle$ of $SR_2$ and $SL_2$ and the eigenstates $|\ell,m
\rangle$ of $L^2$ and $L_z=J_{01}$ are just Clebsch-Gordan
coefficients, which have been worked out once and for all
\cite{landau}. They are real when the relative phases are chosen
by omitting the factor $i$ in (27) and (28) which we will assume
in the following.
\\

With real Clebsch-Gordan coefficients, the inverse expansion of
(\ref{eq:62}) reads:
\begin{equation}\label{eq:62a}
 | sr sl \rangle
 = \sum_{\ell m} (\ell, \frac{k}{2},\frac{k}{2};m,\frac{sr}{2},\frac{sl}{2})|\ell m \rangle
\end{equation}
Inserting the expansion coefficients $ cg^{k\ell}_{sr \;sl}=(\ell,
\frac{k}{2},\frac{k}{2};\frac{sr+sl}{2},\frac{sr}{2},\frac{sl}{2})$
into the expression (\ref{eq:45}) we arrive at the following
expression for the $\psi_{\beta \ell m}$
\begin{equation}\label{eq:63}
\psi_{\beta\ell m}(\alpha,\theta,\phi) = \sum_{ sr}
 N_{\beta \ell} cg^{k 0}_{sr\;-sr}
cg^{k \ell}_{sr\;-sr}e^{i \alpha sr} Y_{\ell m}(\theta,\phi)
\end{equation}
Using the inverse expansion (\ref{eq:62a}), we then find from
(\ref{eq:63}) the following expression for the screw-eigenstates
$\langle \alpha \theta \phi|\beta sr sl\rangle$:
\begin{equation} \label{eq:64}
\psi_{\beta \; sr \; sl}(\alpha, \theta, \phi)
 =\displaystyle \sum_{ \ell sr'} cg^{k\ell}_{sr\;sl}
 N_{\beta \ell} cg^{k 0}_{sr'\;-sr'}
cg^{k \ell}_{sr'\;-sr'}e^{i \alpha sr'} Y_{\ell m}(\theta,\phi)
\end{equation}
where $m=\frac{sr+sl}{2}$.
\\

Our group symmetrical functions can then be expressed using the
equation (\ref{eq:54}). This enable us to determine the
coefficients $a_{s'}^{sr}$ by requiring the resulting function to
be not only invariant to a right-screw along the z-axis, which was
used in  \ref{sec:4} to realise that the admissible $sr$ values
satisfy $sr=0$ mod $N_{\Gamma}$ but also to another specific
Clifford-translation, which together with the translation along
the z-axis spans the group $\Gamma$. For the group $I^*$ we choose
a translation of $\alpha'_{I^*}=\frac{\pi}{5}$ in the direction
$(\theta_{\Gamma},\phi)=(acos(\frac{1}{\sqrt{5}}),0)$ and for the
group $T^*$ a translation of $\alpha'_{T^*}=\frac{\pi}{3}$ in the
direction $(\theta_{\Gamma},\phi)=(acos(-\frac{1}{3}),0)$. For the
group $O^*$ we choose $\alpha'_{O^*}=\frac{\pi}{3}$ in the
direction $(\theta_{\Gamma},\phi)=(acos(\frac{1}{\sqrt{3}}),0)$.
This generalises a technique originally derived for $S^3$/$I^*$ in
\cite{marc rey}, and for all three binary spaces in \cite{marc rey
2}. The invariance means that
\begin{equation}\label{eq:65}
\sum_{sr} a_{s'}^{sr} \sum_{ \ell sr'} cg^{k \ell}_{sr \; sl}
 N_{\beta \ell} cg^{k 0}_{sr' \; -sr'}
cg^{k \ell}_{sr' \; -sr'}(1-e^{i \alpha'_{\Gamma} sr'})e^{i \alpha sr'} Y_{\ell m}(\theta_{\Gamma},0)
\end{equation}
must be identically zero for all $\alpha$. This requires all the
coefficients in the Fourier-expansion in $\alpha$ to vanish, i.e.
\begin{equation}\label{eq:66}
\sum_{sr} C_{sr}^{sr'} a_{s'}^{sr} = 0
\end{equation}
where
\begin{equation}\label{eq:67}
C_{sr}^{sr'} = \sum_{ \ell} cg^{k \ell}_{sr \; sl}
 N_{\beta \ell} cg^{k 0}_{sr' \; -sr'}
cg^{k \ell}_{sr' \; -sr'}(1-e^{i \alpha'_{\Gamma} sr'}) Y_{\ell m}(\theta_{\Gamma},0)
\end{equation}
We can hence determine the $a_{s'}^{sr}$ as the $s'$ eigenvectors with eigenvalue zero, to the matrix
$C^{\dag}C$. Here $sr$ only runs over $sr=0$ mod $N_{\Gamma}$. As the result is independent of $sl$ we can
choose $sl=0$ when doing the computation.
\\

Using (\ref{eq:54}) and (\ref{eq:64}) we get the group symmetrical functions.
\\

We do not in this paper actually use the functions as such, we
only use their expansion for calculating the matrix elements of
the group averaging operator.
\\

\section{The matrix elements $\langle k \ell m|G|k \ell' m' \rangle$}  \label{sec:6}

From the results in the preceding sections we find that the matrix
elements of the group averaging operator are
\begin{eqnarray}\label{eq:68}
& &\langle k \;\ell\; m|G|k\; \ell'\; m' \rangle \nonumber\\
&=& \langle k \;\ell\; m|
(\sum_{sr sl}{|k \;sr \;sl\rangle \langle k \;sr \;sl|})
(\sum_{s'}{|k \;sl \;s'\rangle \langle k \;sl \;s'|})
(\sum_{sr'}{|k \;sr' \;sl\rangle \langle k \;sr' \;sl|})
|k\; \ell'\; m' \rangle \nonumber\\
&=& \sum_{s' \;sl \;sr \;sr'} \langle k\; \ell \; m|k \;sr \;sl \rangle a_{s'}^{sr}\overline{a_{s'}^{sr'}}
\langle k \;sr' \;sl |k \;\ell' \;m' \rangle \nonumber\\
&=& \sum_{s' sl} cg^{k  \ell}_{ 2m-sl \; sl} a_{s'}^{2m-sl}\overline{a_{s'}^{2m'-sl}}cg^{ k \ell'}_{ 2m'-sl \; sl}
\end{eqnarray}
Here the sum over $sl$ should only be extended to values (if any)
that satisfy $2m-sl=2m'-sl=0$ mod $N_{\Gamma}$.

\section{Cosmic expectation value of $C_{\ell}$}  \label{sec:7}

The starting point is the equation (\ref{eq:m11}) in the methods
section, giving the temperature as a function of angle in the sky.
The equation as given, will normally, with complex random Gaussian
variables $X_{\beta s}$, result in a complex temperature signal,
which is due to the fact that we work with complex eigenfunctions
of the Laplacian. As the observed temperature is off course real,
only the real part of the expression should be retained.  Further,
we have to take into account the chosen relative phases of our
wavefunctions $\langle \alpha \theta \phi | \beta \ell m \rangle$,
given in equation (\ref{eq:56}) and (\ref{eq:60a}) of \ref{sec:5}.
Hence we should write
\begin{equation} \label{eq:69}
 2 \delta T(\theta,\phi) =  \sum_{\beta \ell m s} K_{\beta \ell}
  Y_{\ell m}(\phi,\theta) \langle \beta \ell m | \beta s \rangle X_{\beta s} + c.c.
\end{equation}
where $c.c.$ is the complex conjugate of the first expression, i.e. by (\ref{eq:56}) and (\ref{eq:60a})
\begin{equation} \label{eq:70}
 c.c =  - K_{\beta \ell} Y_{\ell -m}(\phi,\theta)(-1)^{\ell-m}
  \langle \beta s | \beta \ell m \rangle \overline{X_{\beta s}}
\end{equation}
Expanding on spherical harmonics
\begin{equation} \label{eq:71}
\delta T(\theta,\phi)= \sum_{\ell m} a_{\ell m} Y_{\ell m}(\phi,\theta)
\end{equation}
We find
\begin{equation} \label{eq:72}
 2 a_{\ell m} = \sum_{\beta s} K_{\beta \ell}
  [\langle \beta \ell m|\beta s \rangle X_{\beta s}
  - (-1)^{\ell-m} \langle \beta s | \beta \ell  -m \rangle \overline{X_{\beta s}}]
\end{equation}
From this we get:
\begin{eqnarray}\label{eq:73}
& &4 a_{\ell m} \overline{a_{\ell m}}  \nonumber\\
&=& \sum_{\beta_1 s_1 \beta_2 s_2} K_{\beta_1 \ell}\overline{K_{\beta_2 \ell}} \nonumber\\
~& &\times[ \langle \beta_1 \ell m|\beta_1 s_1 \rangle X_{\beta_1 s_1}
- (-1)^{\ell-m} \langle \beta_1 s_1 | \beta_1 \ell  -m \rangle \overline{X_{\beta_1 s_1}}] \nonumber\\
~& &\times[ \langle \beta_2 s_2|\beta_2 \ell m \rangle \overline{X_{\beta_2 s_2}}
- (-1)^{\ell-m} \langle \beta_2 \ell  -m | \beta_2 s_2 \rangle X_{\beta_2 s_2}]
\end{eqnarray}
Taking ensemble averages and summing over m, and using (7) in the
methods section gives
\begin{eqnarray}\label{eq:74}
4 \sum_{m}\langle a_{\ell m} \overline{a_{\ell m}}\rangle &=&
\sum_{m \beta s } 2 |K_{\beta \ell}|^2 \langle \beta \ell m |\beta s \rangle \langle \beta s|\beta \ell m \rangle  \nonumber\\
~&=& \sum_{m \beta }2 |K_{\beta \ell}|^2 \langle \beta \ell m |G_{\beta}|\beta \ell m \rangle
\end{eqnarray}
where $G_{\beta}$ is the group averaging operator. Hence we have found the following expression for the
cosmic ensemble average of the $C_{\ell}$:
\begin{equation} \label{eq:75}
\langle C_{\ell} \rangle =\sum_{m}\frac{\langle a_{\ell m} \overline{a_{\ell m}}\rangle}{2\ell+1}
= \frac{1}{2} \sum_{\beta} |K_{\beta \ell}|^2 \sum_{m} \frac{\langle \beta \ell m |G_{\beta}|\beta \ell m \rangle}{2\ell+1}
\end{equation}
We show in \ref{sec:9} that this may be further simplified to
\begin{equation} \label{eq:76}
\langle C_{\ell} \rangle = \frac{1}{2} \sum_{\beta} |K_{\beta \ell}|^2  \frac{multiplicity(\beta)}{\beta^2}
\end{equation}
The significance of this relation, which is stated as a conjecture
in \cite{aurich2}, but in this paper is proven rigorously, is that
it demonstrates that the spectrum can be computed solely from the
radial wave function. The software used (Mathcad) allows to
evaluate this sum up to $\beta$=101. This limit arises from the
fact that Mathcad's symbolic engine, which is used to expand the
Legendre functions (\ref{eq:59}), produces exact rational
coefficients with nominators and denominators that must be smaller
than the largest real which is allowed in Mathcad. It is a
possibility to be explored, that using instead the Fourier
expansion (\ref{eq:45}) would allow this limit to be raised.

\section{Cosmic variance of $C_{\ell}$} \label{sec:8}

We use (\ref{eq:72}) and (\ref{eq:60a})to get

\begin{eqnarray}\label{eq:77}
16 a_{\ell m} \overline{a_{\ell m}} a_{\ell' m'} \overline{a_{\ell' m'}} = \nonumber\\
\sum_{\beta_1 s_1 \beta_2 s_2 \beta_3 s_3 \beta_4 s_4}
  (-1)^{\ell+\ell'}K_{\beta_1 \ell}K_{\beta_2 \ell}K_{\beta_3 \ell'}K_{\beta_4 \ell'} \nonumber\\
~\times [ \langle \beta_1 \ell m|\beta_1 s_1 \rangle X_{\beta_1 s_1}
- (-1)^{\ell-m} \langle \beta_1 s_1 | \beta_1 \ell  -m \rangle \overline{X_{\beta_1 s_1}}] \nonumber\\
~\times [ \langle \beta_2 s_2|\beta_2 \ell m \rangle \overline{X_{\beta_2 s_2}}
- (-1)^{\ell-m} \langle \beta_2 \ell  -m | \beta_2 s_2 \rangle X_{\beta_2 s_2}] \nonumber\\
~\times [ \langle \beta_3 \ell' m'|\beta_3 s_3 \rangle X_{\beta_3 s_3}
- (-1)^{\ell'-m'} \langle \beta_3 s_3 | \beta_3 \ell'  -m' \rangle \overline{X_{\beta_3 s_3}}] \nonumber\\
~\times [ \langle \beta_4 s|\beta_4 \ell' m' \rangle \overline{X_{\beta_4 s_4}}
- (-1)^{\ell'-m'} \langle \beta_4 \ell'  -m' | \beta_4 s_4 \rangle X_{\beta_4 s_4}]
\end{eqnarray}
When we thereafter take the ensemble average, the random variables
only contribute in pairs (see (8) in the methods section). Doing
the pairing between variables in the first and second factor, and
between the third and fourth factor just result in the product $16
\langle a_{\ell m} \overline{a_{\ell m}} \rangle \langle a_{\ell'
m'} \overline{a_{\ell' m'}}\rangle$, while the other possible
pairings produce

\begin{eqnarray}\label{eq:78}
\sum_{\beta_1  \beta_2 }  (-1)^{\ell+\ell'} K_{\beta_1 \ell}K_{\beta_1 \ell'} K_{\beta_2 \ell}K_{\beta_2 \ell'} \nonumber\\
~ \times [- (-1)^{\ell'-m'}\langle \beta_1 \ell m | G_{\beta_1}|\beta_1 \ell'-m' \rangle
   - (-1)^{\ell-m}\langle \beta_1 \ell' m' | G_{\beta_1}|\beta_1 \ell-m \rangle] \nonumber\\
~ \times [- (-1)^{\ell'-m'}\langle \beta_2 \ell' -m' | G_{\beta_2}|\beta_1 \ell m \rangle
   - (-1)^{\ell-m}\langle \beta_2 \ell -m | G_{\beta_2}|\beta_2 \ell' m' \rangle] \nonumber\\
~ +\sum_{\beta_1  \beta_2 } (-1)^{\ell+\ell'} K_{\beta_1 \ell}K_{\beta_1 \ell'} K_{\beta_2 \ell}K_{\beta_2 \ell'}  \nonumber\\
~ \times [\langle \beta_1 \ell m | G_{\beta_1}|\beta_1 \ell' m' \rangle
   + (-1)^{\ell+\ell'-m-m'}\langle \beta_1 \ell' -m' | G_{\beta_1}|\beta_1 \ell-m \rangle] \nonumber\\
~ \times [\langle \beta_2 \ell' m' | G_{\beta_2}|\beta_2 \ell m \rangle
   + (-1)^{\ell+\ell'-m-m'}\langle \beta_2 \ell -m | G_{\beta_2}|\beta_2 \ell' -m' \rangle]
\end{eqnarray}
One should note, that the matrix elements of G are real for all
the manifolds we study in this paper, and hence also symmetric
(this may be shown strictly), a fact we do not exploit here.
Summing over $m$ and $m'$ we realise by substitution of $-m'$ for
$m'$ that the two sums gives identical contributions. Further, the
first factor is the complex conjugate of the second (albeit real),
with $\beta_1$ substituted for $\beta_2$. Our final result becomes
then that the covariance of $C_{\ell}$ is
\begin{equation} \label{eq:79}
Q_{\ell \ell'} =  \langle C_{\ell} C_{\ell'} \rangle -\langle C_{\ell}\rangle \langle C_{\ell'} \rangle
= \frac{1}{2}\frac{1}{2\ell+1}\frac{1}{2\ell'+1} \sum_{m m'} {|M_{m,m'}^{\ell \ell'}|}^2
\end{equation}
where the matrix $M$ is derived by "symmetrizing" the sum of the
matrix elements of the group averaging operator, as follows:
\begin{equation}\label{eq:80}
\fl M_{m,m'}^{\ell \ell'} = \sum_{\beta} K_{\beta \ell}K_{\beta \ell'} \frac{\langle \beta \ell m|G_{\beta}|\beta \ell' m' \rangle
+ (-1)^{\ell+\ell'+m+m'}\langle \beta \ell -m|G_{\beta}|\beta \ell' -m' \rangle}{2}
\end{equation}
We note that all the elements of the covariance matrix are non-negative.
\\

For $S^3$ the matrix $\langle \ell \; m|G_{\beta}| \ell' \; m'
\rangle = \delta_{\ell \;\ell'}\delta_{m \;m'}$ so that the
variance is just
\begin{equation} \label{eq:80a}
Q_{\ell \ell'} =  \delta_{\ell \; \ell'} \frac{2}{2\ell+1}\langle C_{\ell}^2 \rangle
\end{equation}

\section{The formula for the spectrum}  \label{sec:9}

In the \ref{sec:6} we claimed that the formula for the $C_{\ell}$
of equation (\ref{eq:75}) may be simplified to the expression of
equation (\ref{eq:76}).
\\

The starting point is equation (\ref{eq:75}), which may further be
simplified by introducing the projection operator on the
$\ell$-eigenspace $P(\ell) = \sum_m \;|\ell \;m\rangle\langle\ell
\;m\;|$  as
\begin{eqnarray} \label{eq:82}
  (2\ell+1) \langle C_\ell\rangle =\sum_{\beta} \frac{|K_{\beta \ell}|^2}{2}
  \;\mathrm{trace}[G_\beta P(\ell)]
\end{eqnarray}
This, as we demonstrate just below, boils down to
\begin{eqnarray} \label{eq:83}
  \langle C_\ell\rangle = \sum_{\beta > \ell} \frac{|K_{\beta \ell}|^2}{2}
  \;\frac{\mathrm{multiplicity}(\beta)}{\beta^2}
\end{eqnarray}
This equation is very convenient, as it allows us to calculate the
spectrum without calculating the eigenfunctions which are
symmetrical under the holonomy of the manifold considered. The
result holds for all manifolds with holonomies that are
Clifford-translations.
\\

For $S^3$ we get just
\begin{eqnarray} \label{eq:84}
  \langle C_\ell\rangle
  = \sum_{\beta > \ell} \frac{|K_{\beta \ell}|^2}{2}
\end{eqnarray}
The steps leading from equation (\ref{eq:82}) to (\ref{eq:83}) are
based on the observation that both $G$, the group averaging
operator and $P_{\ell}$, the operator that projects on the
eigenspace belonging to the eigenvalue $\ell(\ell+1)$ of the
angular momentum, has certain symmetries.
\\

First we note that $G$ is a sum of right-screw transformations,
and hence commutes with all left-screw transformations, and
therefore also with the generator $SL_2$ of such transformations in
the wz,xy direction:
\begin{equation} \label{eq:85}
SL_2\cdot G-G\cdot SL_2 = 0
\end{equation}
It then follows easily, that the matrix elements of $G$ between
eigenstates of $SR_2$ and $SL_2$ are diagonal in $sl$, because:
\begin{eqnarray} \label{eq:86}
sl \langle sl sr \;| \;G \;| \;sl' sr' \rangle
= \langle sl sr \;|\; SL_2\; \cdot \;G \;|\; sl' sr' \rangle \nonumber \\
= \langle sl sr \;|\; G\; \cdot \; SL_2 \;|\; sl' sr' \rangle
= sl' \langle sl sr \;|\; G \;|\; sl' sr'\rangle
\end{eqnarray}
and therefore:
\begin{equation} \label{eq:87}
\langle sl \;sr \;|\; G \;|\; sl' sr'\rangle
= \langle sl \;sr \;|\; G \;|\; sl \;sr'\rangle \delta_{sl sl'}
\end{equation}
Further, the operator $AL_2$ for lifting $sl$ also commutes with
$G$ from which we by standard arguments can conclude that the
diagonal elements are independent of $sl$. By construction:
\begin{eqnarray} \label{eq:88}
AL_2\;|\; sl \;sr\rangle = al(sl)\;|\;sl+2\;sr\rangle \qquad
\langle sl \;sr\;|\;AL_2^{\dag} = \overline{al(sl)}\; \langle sl+2\; sr\;|
\end{eqnarray}
Noting that $AL_2^{\dag}$ acting on the state $|\;sl+2 sr \rangle$
is proportional to $|\;sl \;sr \rangle$ and that, by complex
conjugation
\begin{equation} \label{eq:89}
\langle sl+2 \;sr \;|\; AL_2 \;|\; sl sr\rangle = al(sl) \qquad
\langle sl \;sr \;|\; AL_2^{\dag} \;|\; sl+2 \;sr\rangle = \overline{al(sl)}
\end{equation}
we see that
\begin{equation} \label{eq:90}
 AL_2^{\dag} \;|\; sl+2 \;sr\rangle = \overline{al(sl)} \;|\; sl sr\rangle \qquad
\langle s1+2 \;sr\;|\;AL_2 = al(sl) \langle s1 \;sr\;|
\end{equation}
From (\ref{eq:88}) and (\ref{eq:89}) follows
\begin{eqnarray} \label{eq:91}
&&al(sl) \langle sl+2 \;sr \;|\; G \;|\; sl+2 \;sr\rangle
= \langle sl+2 \; sr \; | \; G \; \cdot \; AL_2 \; | \; sl+2 \; sr \rangle \nonumber\\
&&= \langle sl+2 \; sr \; | \; AL_2 \; \cdot \; G \; | \; sl+2 \; sr \rangle
= al(sl) \langle sl \;sr \; | \; G \; | \; sl \; sr \rangle
\end{eqnarray}
demonstrating that the diagonal matrix elements of $G$ are
independent of $sl$.
\\

As the trace of $G$ is equal to the dimension of the space of
non-vanishing symmetrical states, i.e. the multiplicity, we find
that, for each of the $\beta$ values of $sl$:
\begin{equation} \label{eq:92}
\sum_{sr }\langle sl \;sr \;|\; G \;|\; sl \;sr\rangle
=\frac{1}{\beta} \sum_{sl sr}\langle sl \;sr \;|\; G \;|\; sl \;sr\rangle
= \frac{multiplicity(\beta)}{\beta}
\end{equation}
For the operator $P_{\ell}$, we find, firstly, that the only
nonzero matrix elements are the ones satisfying the selection rule
$sl+sr = sl'+sr'$ and secondly, although not so easily, that the
sum
\begin{equation} \label{eq:93}
\sum_{sl} \langle sr \;sl \;|\;P_{\ell}\;|\;sr \;sl \rangle
\end{equation}
is independent of $sr$ and hence for each of the $\beta$ values of
$sr$ must be:
\begin{equation} \label{eq:94}
\sum_{sl} \langle sr \;sl \;|\;P_{\ell}\;|\;sr \;sl \rangle =
\frac{1}{\beta} \sum_{sl sr} \langle sr \;sl \;|\;P_{\ell}\;|\;sr \;sl \rangle
= \frac{2 \ell +1}{\beta}
\end{equation}
Combining the two equations, we realise that
\begin{eqnarray} \label{eq:95}
\mathrm{trace}(G P_{l}) &=& \sum_{sl sr sl' sr'} \langle sl \;sr \;|\;G\;|\;sl' sr' \rangle
\langle sl' sr'\;|\;P_{\ell}\;|\;sl \;sr \rangle \nonumber \\
&=& \;\;\; \sum_{sl sr} \langle sl \;sr \;|\;G\;|\;sl \;sr \rangle
\langle sl \;sr\;|\;P_{\ell}\;|\;sl \;sr \rangle
\end{eqnarray}
because equation (77) requires $sl=sl'$ and the selection rule for
$P_{\ell}$ then $sr=sr'$ , which by the independence of the $G$
term on $sl$ allow us to write
\begin{eqnarray} \label{eq:96}
& &\mathrm{trace}(G P_{l})
=\sum_{sr} \langle sl \;sr \;|\;G\;|\;sl \;sr \rangle
\sum_{sl} \langle sl \;sr\;|\;P_{\ell}\;|\;sl \;sr \rangle \nonumber\\
&=& \frac{multiplicity(\beta)}{\beta} \frac{2 \ell +1}{\beta}
\end{eqnarray}
which finishes our proof, apart from proving the above stated
properties of $P_{\ell}$.
\\

To do that, first note that $SR_2+SL_2$ is just $2*J_{01}$ and
hence commutes with $P_{\ell}$. Therefore
\begin{equation} \label{eq:97}
(sl+sr) \langle sl\;sr\;|\;P_{\ell}\;|\;sl'\;sr' \rangle
=(sl'+sr') \langle sl\;sr\;|\;P_{\ell}\;|\;sl'\;sr' \rangle
\end{equation}
showing that if $(sl+sr) \ne (sl'+sr')$ the matrix element
vanishes.
\\

The further remaining detail, of showing independence of
$\sum_{sl} \langle sr \;sl \;|\;P_{\ell}\;|\;sr sl \rangle$ of
$sr$ is accomplished by noting that the operator $AL_2+AR_2$ is
just $2*(J_{0 1}+ i J_{2 0})$ and hence commute with $P_{\ell}$.
\\

We use this to evaluate
\begin{equation} \label{eq:98}
\langle sl\;sr+2\;|\;P_{\ell}\;(AL_2+AR_2)\;|\;sl\;sr\rangle
= \langle sl\;sr+2\;|\;(AL_2+AR_2)\;P_{\ell}\;|\;sl+2\;sr\rangle
\end{equation}
Noting that we have equations for the action of $AR_2$ analogous
to the above for $AL_2$, we find, by letting $AL_2+AR_2$ act to
the right on the ket in the first expression, and to the left on
the bra in the second
\begin{eqnarray} \label{eq:99}
al(sl)\;\langle sl\;sr+2\;|\;P_{\ell}\;|\;sl+2\;sr\rangle
+ar(sr)\;\langle sl\;sr+2\;|\;P_{\ell}\;|\;sl\;sr+2\rangle \nonumber \\
= al(sl-2)\;\langle sl-2\;sr+2\;|\;P_{\ell}\;|\;sl\;sr\rangle
+ar(sr)\;\langle sl\;sr\;|\;P_{\ell}\;|\;sl\;sr\rangle
\end{eqnarray}
Next we sum over $sl$, and note that in doing that, we can replace
the $sl-2$ in the first term on the right hand side of the
equation with $sl$ (we can do that, because $al(sl)$ is zero when
$sl$ is the maximum value $sl_{max}=\beta-1$). The sum of this
term then equals the sum of the first term on the left hand side,
and both sums can be neglected. We are left with the sums over the
second terms
\begin{equation} \label{eq:100}
\sum_{sl} ar(sr)\langle sl\;sr+2\;|\;P_{\ell}\;|\;sl\;sr+2\rangle
=\sum_{sl} ar(sr)\langle sl\;sr\;|\;P_{\ell}\;|\;sl\;sr\rangle
\end{equation}
which demonstrates that the sum is independent of $sr$ which
completes the proof.

\section{Likelihood function} \label{sec:10}

Central to extracting uncertainty-bounds for the cosmological
parameters from the observed properties of the CMB is the setting
up of a likelihood function. In the case of flat space, or $S^3$
models, this is relatively straightforward, as the $a_{\ell m}$'s
can be considered to be independently distributed random Gaussian
variables, so that the $C_{\ell}$'s have a chi-square distribution
with $2\ell+1$ degrees of freedom.
\\

In the case of a nontrivial global topology for the universe, as
the $S^3/I^*$, $S^3/O^*$ and $S^3/T^*$ studied here, the
situation is more complicated.
\\

The $C_\ell$'s are then definitely not chi-square distributed, and
the $C_{\ell}$'s are not independent.
\\

One may in principle calculate the likelihood distributions from
the assumed Gaussian distributions of the random amplitudes $X$ by
a large number of simulations, for each choice of the cosmological
parameters. Such grid-based approaches are very demanding in terms
of computer-time.
\\

The strategy used by the WMAP team, for nearly flat models, is
instead to sample the cosmological parameter-space by setting up
Markov chain Monte-Carlo simulations \cite{verde}. That approach
is critically dependent on the ability to specify the conditional
probability distribution for the observations, given the model,
which is a simple task when the $C_{\ell}$'s are
chi-square-distributed, but not feasible in the case of
$S^3$/$\Gamma$.
\\

In this preliminary analysis it was chosen to use a crude
approximation instead, by assuming the $C_{\ell}$'s
to have the Gaussian distribution of (\ref{eq:104}).
\\

From the ensemble-averages of  \ref{sec:7} and \ref{sec:8} for the
$C_{\ell}$'s and their covariance, such a multidimensional
Gaussian distribution is easily constructed from the covariance
matrix $Q$ of (\ref{eq:79}):
\begin{equation} \label{eq:104}
W(C_{\ell}^{obs} | model) = \frac{1}{\sqrt{det(2\pi Q^{th})}}
e^{-\frac{1}{2}\sum_{\ell,\ell'} (C_{\ell}^{obs} -C_{\ell}^{th}) {{Q^{th}}^{-1}}_{\ell,\ell'}(C_{\ell'}^{obs} -C_{\ell'}^{th})}
\end{equation}
Here the probability depends on the model parameters through the
dependence of $C_{\ell}^{th}$ and ${Q^{th}}$ on the "theory", i.e.
the model parameters. In a more refined analysis, one has to
consider the variance due to measurement uncertainties, but for
the low-$\ell$ power, which we consider here, the cosmological
variance dominates. Hence we ignore the complications of
measurement uncertainty in the present analysis.
\\

Often a prior is applied, i.e. an apriori distribution
$P^{ante}(model) $ for the model parameters, which may be as
simple as to constrain the parameters to a certain window in
parameter space.  In any case, application of Bayes principle
result in the following likelihood function for the model
parameters, given the observed $C_{\ell}$'s:
\begin{equation} \label{eq:105}
 \mathcal{L}(model | C_{\ell}^{obs})
 = \frac{W(C_{\ell}^{obs}|model)*P^{ante}(model)}
 {\sum_{model}{W(C_{\ell}^{obs}|model)*P^{ante}(model)}}
\end{equation}
Two different apriori distributions are used in this paper: 
a uniform primer over the cosmic window studied, and a 
Gaussian primer with means and standard deviations corresponding to the WMAP estimate for the pl-model, 
$\Omega_{mass} = 0.257 +/- .025$ and the general estimate \cite{spergel} of
$\Omega_{tot} = 1.02 +/- 0.02$ .
\\

It's a valid question to ask, how good the ansatz (\ref{eq:104})
is. It's clearly unphysical, in as much as the $C_{\ell}$'s are
inherently positive. It's also known to be slightly biased
\cite{verde}, in the case of $S^3$ where the exact distribution is
\begin{equation} \label{eq:106}
W^{S^3}(C_{\ell}^{obs} | model) = \prod_{\ell}{\chi(C_{\ell}^{obs},C_{\ell}^{th},Q^{th}_{\ell\ell})}
\end{equation}
i.e. a product of chi-square distributions with mean $\langle
C_{\ell}^{obs} \rangle = C_{\ell}^{th}$ and $2\ell+1 = 2 \;
{C_{\ell}^{th}}^2 / Q^{th}_{\ell\ell} $ degrees of freedom.
However, the distribution (\ref{eq:104}) has the proper mean
values, if we neglect that the $C_{\ell}$'s
are inherently positive ,for all the manifolds $S^3$/$\Gamma$, derivable from the
theoretical relations of the preceding sections:
\begin{eqnarray} \label{eq:107}
 \langle C_{\ell}^{obs} \rangle =  C_{\ell}^{th} \nonumber\\
 \langle C_{\ell}^{obs} C_{\ell'}^{obs} \rangle
  - \langle C_{\ell}^{obs} \rangle \langle C_{\ell'}^{obs} \rangle
  = Q^{th}_{\ell \ell'} \nonumber\\
 \langle R^2 \rangle = \langle \sum_{\ell,\ell'} {(C_{\ell}^{obs} -C_{\ell}^{th})
   {{Q^{th}}^{-1}}_{\ell,\ell'}(C_{\ell'}^{obs} -C_{\ell'}^{th})}
\rangle = 14
\end{eqnarray}
where the sum over $\ell$ and $\ell'$ runs from 2 to 15.
\\

Doing a direct simulation of (\ref{eq:72}), for an ensemble of
10.000 Universes, which takes about 20 minutes calculation time
for each choice of cosmological parameters, the resulting
distributions for the $C_{\ell}$'s can be found, and they are
reasonably approximated by chi-square distributions, with an
effective degree of freedom of $D = 2 \;{C_{\ell}^{th}}^2 \; /\;
Q^{th}_{\ell\ell}$, with some exceptions, however, as seen in
Figure \ref{fig:a3}. As an alternative to (\ref{eq:104}) the
following 2 probability functions have been tested:
\begin{equation} \label{eq:107a}
W^{\Gamma}(C_{\ell}^{obs} | model) = \prod_{\ell}{\chi(C_{\ell}^{obs},C_{\ell}^{th},Q^{th}_{\ell\ell})}
\end{equation}
which means that the off-diagonal elements of $Q^{th}$ are simply
neglected, as well as the distribution:
\begin{equation} \label{eq:107b}
W^{\Gamma}(C_{\ell}^{obs} | model) = det{P}
\prod_{\ell}{ \chi(\widetilde{C}_{\ell}^{obs}, \widetilde{C}_{\ell}^{th}, 1)}
= det{P}
\prod_{\ell}{ \chi(P \cdot C_{\ell}^{obs},P \cdot C_{\ell}^{th}, 1)}
\end{equation}
where the matrix $P$ is used to get a unit covariance matrix:
\begin{equation} \label{eq:107c}
\widetilde{C}_{\ell}^{obs} = P \cdot C_{\ell}^{obs} \qquad
\widetilde{C}_{\ell}^{th} = P \cdot C_{\ell}^{th} \qquad
P' \cdot P =  {Q^{th}}^{-1}
\end{equation}
The function (\ref{eq:107b}) has the proper mean values
(\ref{eq:107}) whereas this is not the case for the function
(\ref{eq:107a}).
\\

\begin{figure}
\centerline{\includegraphics[width=15cm]{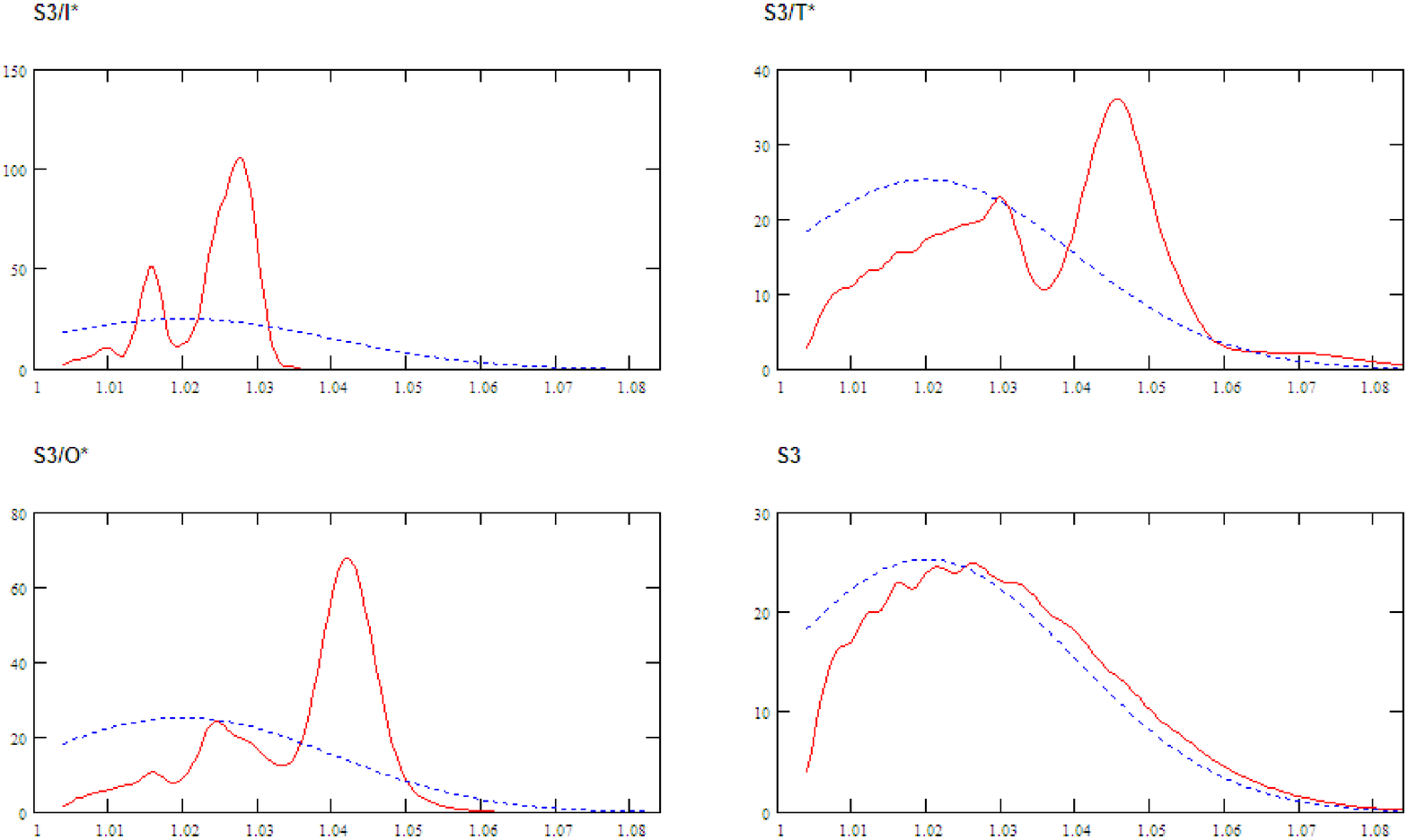}}
\caption{The likelihood distribution as a function of
$\Omega_{tot}$ along the most likely value of $\Omega_{mass}$ =
0.26 for $S^3/O^*$, $S^3/T^*$, $S^3$ and = 0.25 for $S^3/I^*$,
with the probability function (\ref{eq:107a}). Compare with Figure
\ref{Figure2} of  section \ref{sec:results} . Dotted line shows
the prior distribution, for comparison. Note the different scales
on the vertical axes.} \label{fig:a1}
\end{figure}

\begin{figure}
\centerline{\includegraphics[width=15cm]{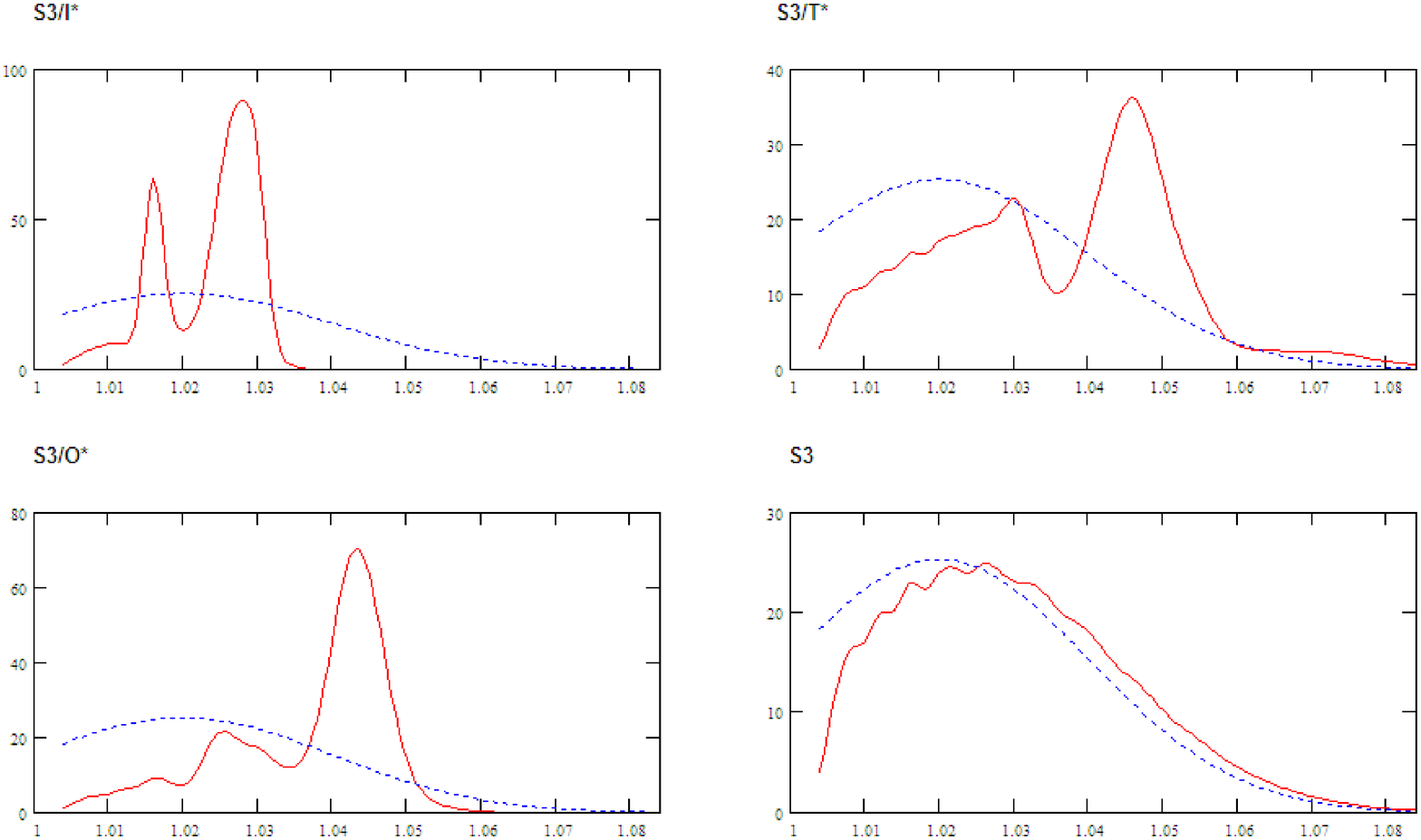}}
\caption{The likelihood distribution as a function of
$\Omega_{tot}$ along the most likely value of $\Omega_{mass}$ =
0.26 for $S^3/O^*$, $S^3/T^*$, $S^3$ and $S^3/I^*$, with the
probability function (\ref{eq:107b}). Compare with Figure
\ref{Figure2} of section \ref{sec:results} . Dotted line shows the
prior distribution, for comparison. Note the different scales on
the vertical axes.} \label{fig:a2}
\end{figure}

\begin{figure}
\centerline{\includegraphics[width=15cm]{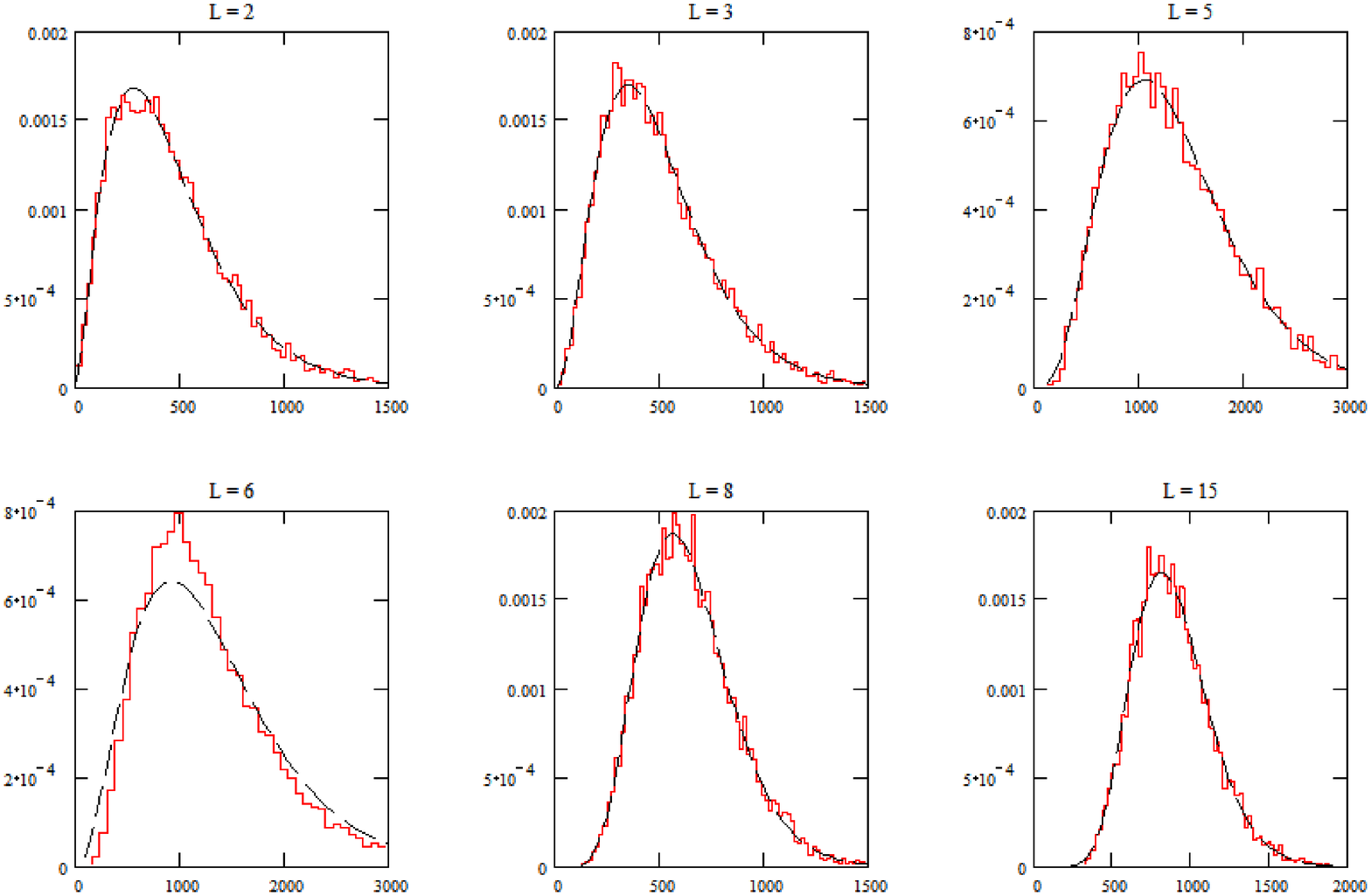}}
\caption{The simulated histograms for $S^3/I^*$ of the moments
$\ell(\ell+1)*C_{\ell}/2\pi$ for selected $\ell$-values
(calculated for $\Omega_{mass}$ = 0.26 and $\Omega_{tot}$ =
1.028). \label{fig:a3} The dashed lines show chi-square
distributions with the same mean and variance as the histograms.
The simulated distributions are clearly not chi-square, especially
for $\ell=6$ and possibly for $\ell=2$.}
\end{figure}

Figure \ref{fig:a1} and Figure \ref{fig:a2} illustrate, that the
choice of the probability function $W$ has an impact on the
calculated ex post likelihood surfaces. Using (\ref{eq:107a}) or
(\ref{eq:107b}) rather than (\ref{eq:106}) has f.x. the effect,
that the secondary optimum for $S^3/I^*$ near $\Omega_{tot}=1.017$
becomes somewhat larger, but still smaller, than the peak near
$\Omega_{tot}=1.030$. Also the peak at $1.03$ shifts slightly
towards lower values of $\Omega_{tot}$. Hence, the issue of
establishing a proper probability function $W$ remains an
important question for examining the best fit estimates for the
cosmological parameters in case of the nontrivial topologies
$S^3/\Gamma$.
\\

\begin{figure}
\centerline{\includegraphics[width=10cm]{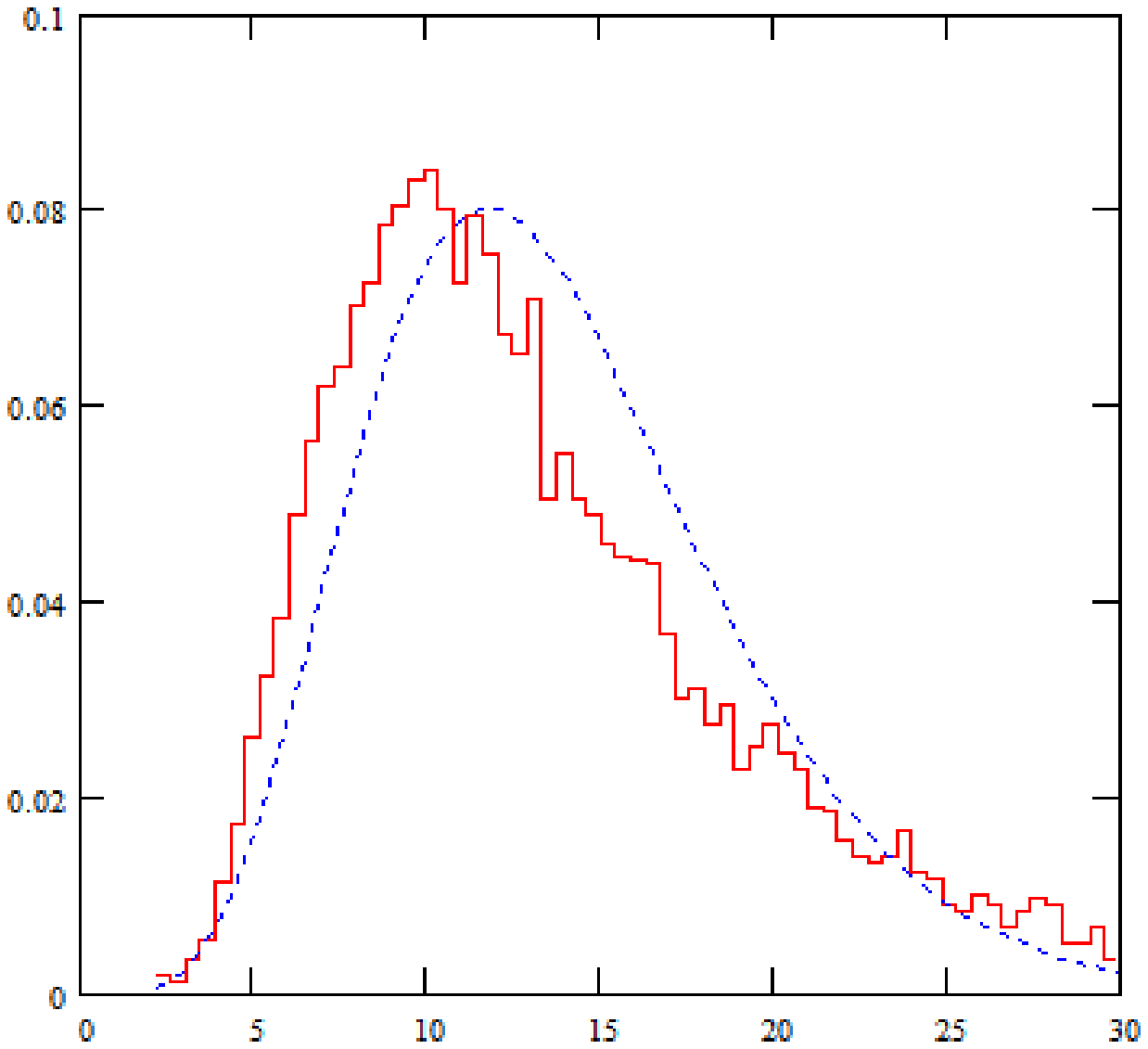}} \caption{The
simulated histogram for $S^3/I^*$ of the $R^2$ defined by
(\ref{eq:107}), calculated for $\Omega_{mass}$ = 0.26 and
$\Omega_{tot}$ = 1.028 and 10.000 universes. Horizontal axis is
$R^2$. The simulated distribution is not chi-square, with 14
degrees of freedom, showing that the ansatz (\ref{eq:104}) is not
perfect.} \label{fig:a3_2}
\end{figure}

Neither of the proposed distributions $W$ are exact. Hence the
proper solution in future work could be to use that the $a_{\ell
m}$'s do in fact have a multidimensional Gaussian distribution,
which can be expressed by the covariance matrix $Q_{\ell m \ell'
m'} = \langle a_{\ell m} a_{\ell' m'} \rangle$. The input to the
likelihood calculation would then need to be the observed $a_{\ell
m}$'s instead of the observed $C_{\ell}$'s. This approach would be
complicated by the anisotropy of the $a_{\ell m}$'s which are
found for all the non-trivial topologies $S^3/\Gamma$, as the
theoretical ensemble averages of $\langle a_{\ell m} a_{\ell' m'}
\rangle$ depend on the alignment of the special directions. If
these directions can be found by a matching circle's detection,
the procedure could be used quite straightforwardly, however, to
get an improved estimate of the cosmological parameters. Another
intriguing possibility, in case the matched circles search does
not produce a definite answer, could be that a systematic rotation
in the 3-parameter space of rotations on the sky of the $a_{\ell
m}$'s, although computationally challenging, could yield a best
fit alignment between observations and theory, thus identifying
any preferred cosmic directions, even without recourse to finding
matching circles.
\\

It's rather easy to derive, that (\ref{eq:104}) implies that the
last quantity $R^2$ is distributed as $\chi(R^2,14,14)$, for all
the manifolds $S^3$/$\Gamma$. This may be compared with the exact
distribution of $R^2$ for $S^3/I^*$ derivable by simulation from
(\ref{eq:72}), shown in Figure \ref{fig:a3_2}.
\\

It is seen that such a chi-square distribution does not  reproduce
the simulated distribution of $R^2$ fairly well, for the case of
$S^3/I^*$. For the more general manifolds such as $S^3$/$\Gamma$
we do not, however, have an explicit, exact expression for the
true multidimensional distribution $W(C_{\ell}^{obs} | model)$
(but it could be simulated by drawing a large number of the random
variables $X_{\beta,s}$).
\\

Nevertheless, as the distribution (\ref{eq:104}) does reproduce
the mean values (\ref{eq:107}) if we can neglect that the $C_\ell$'s
are positive, we will assume that
(\ref{eq:104}) is a workable approximation, at this preliminary
state of analysis.
\\

A significant lesson from Figure \ref{fig:a3_2} is, that the
statistics $R^2$, which is the sum of the squared deviation of the
observed $C_\ell$'s from their cosmological ensemble averages,
measured relative to the expected variance, is a fairly broad
distribution. As it is further found, that the observed value of
$R^2$ is fairly close to its expected value of 14, for all
topologies and across the cosmological parameter space, we can not
use $R^2$ as a statistic to reject any specific topology, or any
choice of the cosmological parameters.
\\

Whereas the distribution (\ref{eq:104}) is very sensitive to
deviations of $C_{\ell}^{obs}$ from $C_{\ell}^{th}$ this is not
the case for the distribution of $R^2$ because of the "phase space
factor", i.e. the fact that the surface to radius ratio in the
14-dimensional space of the $C_{\ell}$'s grows quickly with $R$.
This means, that we can use the likelihood distribution
(\ref{eq:104}) to pinpoint the best values of the cosmological
parameters, given the topology, but cannot easily use it
to discriminate between alternative topologies.
\\

\section{2-point angular correlation function} \label{sec:11}

The 2-point angular correlation function $C(\theta)$ is defined as
the average over the sky of $C(\theta) = av(\delta T(\hat{n})
\delta T(\hat{n}'))$ with $\hat{n}\cdot\hat{n}'=cos(\theta)$,
which is related to the moments $C_{\ell}$ by
\begin{equation}
C(\theta) = \frac{1}{4\pi}\sum_{\ell=2}^{\infty}{(2\ell+1)C_{\ell} P_{\ell}(cos(\theta))}
\end{equation}
where $P_{\ell}$ is the Legendre function. The cosmic expectation
value of $C(\theta)$ is found by replacing $C_{\ell}$ with
$\langle C_{\ell} \rangle$ and similarly, the cosmic variance of
$C(\theta)$ is
\begin{equation}
 \langle {C(\theta)}^2 \rangle - {\langle C(\theta) \rangle}^2
 = \frac{1}{(4\pi)^2}\sum_{\ell=2, \ell'=2}^{\infty}
 {(2\ell+1)(2\ell'+1)Q_{\ell,\ell'} P_{\ell}(cos(\theta))P_{\ell'}(cos(\theta))}
\end{equation}
As discussed in \cite{spergel} the observed two-point correlation
function is very flat for large angles, a feature which the
WMAP-models based on flat space or nearly flat space are unable to
reproduce. The anomaly is mainly a result of the very low observed
quadrupole and octopole moments. To quantify the anomaly,
simulations are reported of the statistic
\begin{equation}
S(\rho) = \int_{-1}^{cos(\rho)} {C(\theta)}^2 dcos(\theta)
\end{equation}
It is found, for the best fit $\Lambda CDM$ model, for $\rho$ = 60
degrees that only 0.15\% of the simulations have lower value than
the observed value of $S(\rho)$. For the running index model, the
similar result of simulations is found to be 0.3\%.
\\

\begin{figure}
\centerline{\includegraphics[width=10cm]{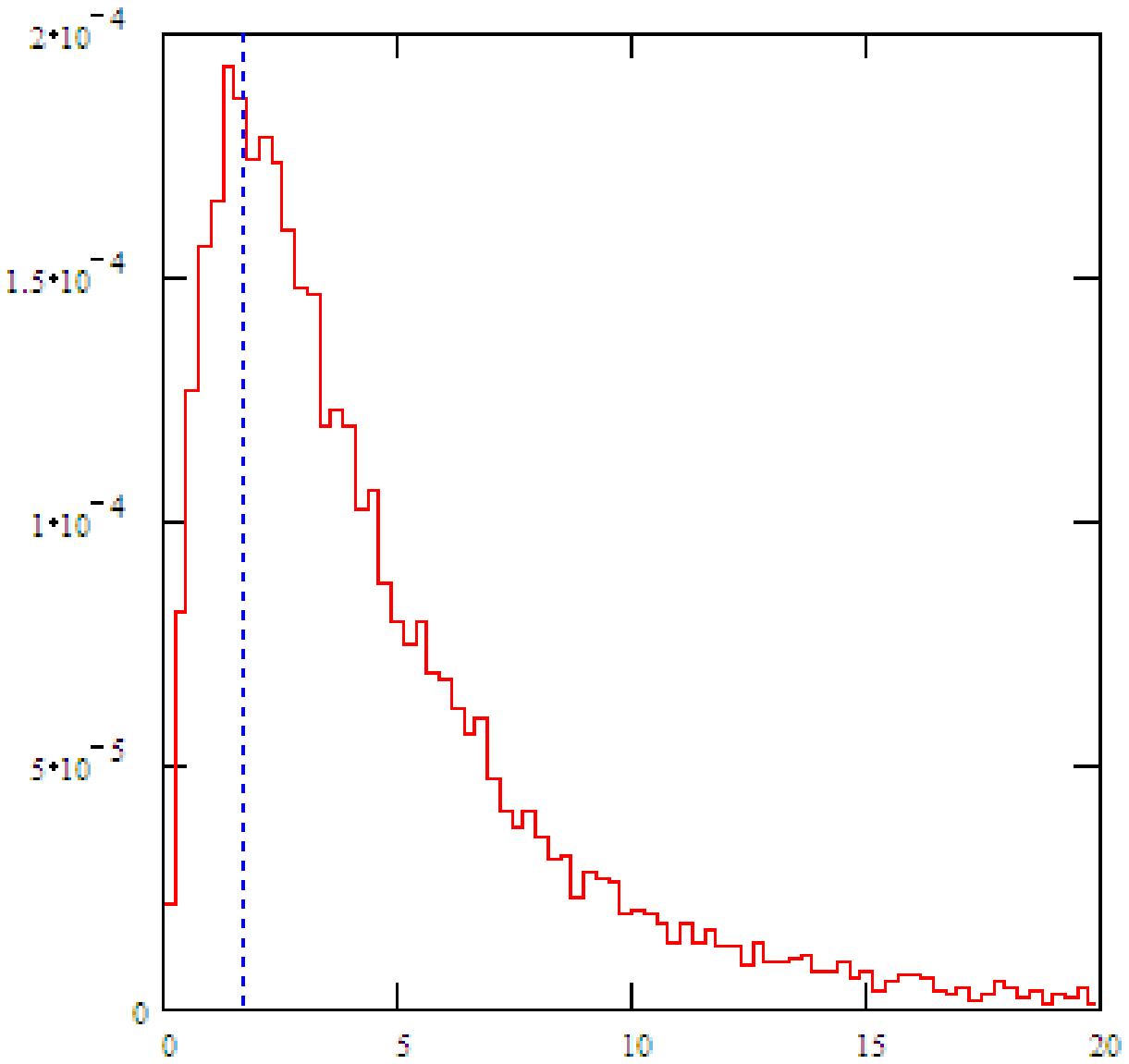}} \caption{The
simulated histogram for $S^3/I^*$ of the $S(60)$ statistic,
calculated for $\Omega_{mass}$ = 0.26 and $\Omega_{tot}$ = 1.028
and 10.000 universes. Horizontal axis is $S(60)$ times
${10}^{-3}$. The vertical line shows the observed value. The
probability of getting a value less than the observed is 23\%. The
simulated distribution is not chi-square, but disregarding this
one finds the effective degrees of freedom $2 {mean}^2/{varianse}$
to be 2.26, showing that the statistic is sampling very few random
variables.} \label{fig:a4}
\end{figure}

In this paper, we only study the S-statistic filtered to a maximum
value of $\ell$ of 15, but the results are quite similar as the
S-statistic heavily emphasises the lowest multipoles. As an
example, it is found that for $\Omega_{mass}$ = 0.26 and
$\Omega_{tot}$ = 1.028 (the favoured value from this study) the
$S^3$ model with a power law spectrum, gives an $S(60)$ lower than
the observed value only in 0.13\% of simulations. This is in
contrast with simulations for $S^3/I^*$ where the simulations
performed, for the same values of the cosmological parameters,
give a 23\% chance of a lower $S(60)$ than observed, see Figure
\ref{fig:a4}.

In \cite{aurich2} the cosmological expectation value of the
S-statistic is used to locate the values of the cosmological
parameters that brings this expectation value closest to the
observed value.
\\

This might be problematic, however, because the choice of
statistics is heavily biased to emphasize the low values of the
two first multipole moments seen in the observations, as discussed
in \cite{efstathiou}. For that reason, it was preferred in this
paper, rather to use the maximum likelihood principle, even though
it is based on only an approximate distribution $W$, to search for
the most likely values in cosmological parameter space for each of
the topologies. As shown in Figure \ref{Figure1} and Figure
\ref{Figure2} the maximum likelihood is found for
$\Omega_{mass}$=0.26 and $\Omega_{tot}$ = 1.028 +/- 0.0023. A
secondary but smaller peak, however, is found near the value of
$\Omega_{tot}$ found in \cite{aurich2} by minimising the
S-statistic. The center value and width of this peak yields
$\Omega_{tot}$ = 1.017 +/- .0015. As shown in Figure \ref{fig:a1}
and Figure \ref{fig:a2} this secondary peak becomes somewhat more
pronounced when using the alternative approximate distributions
(\ref{eq:107a}) or (\ref{eq:107b}).

\end{document}